\documentclass[journal]{IEEEtran}

\usepackage{amssymb}
\usepackage[cmex10]{amsmath}
\usepackage{graphicx}
\usepackage{epstopdf}
\usepackage{array}
\usepackage{mathabx}
\usepackage{multirow}
\usepackage[caption=false,font=footnotesize]{subfig}
\usepackage{cite}

\newcounter{MYtempeqncnt}

\begin{document}

\title{Equalized Time Reversal Beamforming for Indoor Wireless Communications}

\author{Carlos~A.~Viteri-Mera, \IEEEmembership{Student Member, IEEE}, and Fernando~L.~Teixeira, \IEEEmembership{Fellow, IEEE}% <-this % stops a space
\thanks{This work has been supported in part by the Ohio Supercomputer Center (OSC) under grants PAS-0061 and PAS-0110, and by a Fulbright Colombia Fellowship.}
\thanks{The authors are with the ElectroScience Laboratory, Department of Electrical and Computer Engineering, The Ohio State University, 1330 Kinnear Rd., Columbus, OH 43212 USA, (614) 292-6993 (e-mail: \{viteri.5,teixeira.5\}@osu.edu).}%
\thanks{C. Viteri-Mera is also with the Department of Electronics Engineering, Univeridad de Nari\~no, Pasto, Colombia.}}%

\maketitle

\begin{abstract}
Time-reversal (TR) is a beamforming technique for frequency-selective channels, which has received increasing attention due to its high energy efficiency and low computational complexity for wireless communications. In this paper, we present two contributions on time-reversal (TR) wireless beamforming for single-user indoor wideband MISO systems. First, we provide novel analyses of a baseband TR system using two commonly used indoor propagation channel models. We derive closed-form approximations for the inter-symbol interference (ISI) with these channel models in order to characterize the influence of propagation conditions (such as the power-delay profile, delay spread, and bandwidth) on TR performance metrics. In particular, we analyze spatial focusing and time compression performance of TR beamforming, and their impact on the bit error rate (BER). As a second contribution, we introduce an equalized TR (ETR) technique that mitigates the ISI of conventional TR. The proposed ETR utilizes a zero-forcing pre-equalizer at the transmitter in a cascade configuration with the TR pre-filter. Unlike previous approaches to ISI mitigation in TR systems, we derive theoretical performance bounds for ETR and show that it greatly enhances the BER performance of conventional TR with minimal impact to its beamforming capabilities. By means of numerical simulations, we verify our closed-form approximations and show that the proposed ETR technique outperforms conventional TR with respect to the BER under any SNR.
\end{abstract}

\section{Introduction}

Very short-range wireless architectures, such as pico and femtocells, are becoming ubiquitous as data volume increases and spectrum scarcity makes high-density deployments more feasible economically \cite{bernhard2010}. Short-range solutions can be used to offload cellular network traffic to wireless local area networks (WLAN), as seen with the proliferation of indoor WiFi hotspots. Because of their smaller size and increasing operating frequencies, these architectures, as well as future types of indoor networks, may adopt access points (AP) that employ irregularly-spaced or other unconventional antenna arrays instead of the arrays in use today. New beamforming techniques that perform well in such scenarios are thus highly desirable\footnote{The term beamforming is traditionally used to denote phased array techniques for beam-steering in flat-fading channels, i.e. operating in the 2D manifold spanned by the azimuth and elevation angles. In this paper, we use the term beamforming in a broader sense to denote signal processing techniques for frequency-selective multipath channels, that allow spatial focusing of RF power in co-range as well (3D), or even in time (4D space-time beamforming) \cite{yavuz2009,oestges2005,elsallabi2010}.}.

One of the techniques with potential to provide advanced beamforming capabilities in rich scattering scenarios is time-reversal (TR) \cite{fouda2012}. Considering the radio channel as a linear system \cite{fink1997}, TR is a signal transmission technique that uses the time-reversed channel impulse response (CIR) as a linear filter applied to the transmitted signal. Such pre-filter enables spatial focusing of the signal at the receiver and compression of the CIR in the time domain \cite{fouda2012c, oestges2005, nguyen2006, elsallabi2010, han2016}. First, spatial focusing in TR occurs because all multipath components add coherently at the receiver's location, while they combine incoherently in other positions in space; this is allowed by the spatial signature contained in the CIR. Second, in-phase addition of multipath components takes place at specific sampling instants. This effect is due to the matched filter behavior of the TR pre-filter, which also has partial equalization properties that reduce inter-symbol interference (ISI) \cite{kyritsi2004}. Due to this appealing characteristics, TR beamforming is particularly attractive for indoor pico and femtocells, where the channel is typically slow-varying and rich scattering is prevalent. In such scenarios, spatial focusing can be maintained without requiring a fast update of the channel state information (CSI). In addition, the main advantage of TR with respect to conventional multi-carrier systems in use today is the reduced computational complexity at the transmitter and, specially, at the receiver \cite{chen2013}. General advantages stemming from beamforming towards green wireless systems~\cite{cardoso2013,goncalves2012} also exist, with TR receiving special attention for its potential use to improve energy efficiency in future wireless networks \cite{wang2011,chen2011}.

Because of their high temporal resolution, most of the work in TR has focused on ultrawideband (UWB) systems, although the suitability of this technique in conventional wideband systems has been verified as well \cite{kyritsi2004, nguyen2006}. The performance of TR, in terms of bit error rate (BER) and focusing capability, has been addressed by means of empirical and theoretical approaches. In \cite{oestges2005}, the authors study the space-time focusing of a single-input single-output (SISO) TR system in two scattering scenarios; they define performance metrics and find empirical formulas for them. References \cite{wang2011} and \cite{han2012} present a theoretical analysis on space-time focusing under single user SISO and multi-user multiple-input single-output (MISO) systems, respectively. The probability of bit error in TR systems has been investigated both theoretically \cite{nguyen2006,wang2011,han2012}, and empirically (BER) \cite{kyritsi2005}. These works focus on separating the received signal components into desired signal power and ISI power (inter-user interference is also characterized in some cases) in order to obtain approximations to the signal-to-interference-plus-noise-ratio (SINR). However, the error of these approximations and their sensitivity to propagation conditions have not been analyzed. For example, the influence of the channel power-delay profile (PDP) on the TR beamforming system is unknown. 

This motivates the first contribution of this paper, which is the performance characterization of conventional TR beamforming for single-user indoor MISO channels in typical pico and femtocells. Our analysis is based on two statistical channel models \cite{saleh1987} with different PDPs that are well suited for such scenarios. We use these two indoor channel models to provide a novel performance comparison of TR beamforming techniques under different propagation conditions, viz. delay spread, sampling time (bandwidth), and CIR duration. We derive closed-form approximations to the probability of bit error and space-time focusing performance parameters. We find that performance is highly dependent on the propagation conditions and, hence, the relevance of the presented analysis.

In the second part of this paper, we propose an equalized TR technique based on a previous contribution by our group \cite{fouda2012}. We focus on a single-user MISO frequency-selective channel scenario, operating in conventional wideband systems with low complexity receivers. A number of works have addressed the problem of mitigating ISI in TR. For example, in \cite{kyritsi2005} the authors propose the joint design of a TR pre-filter and a zero-forcing (ZF) pre-equalizer by finding the pre-filter closest to TR that sets the ISI power to zero. A similar approach is presented in \cite{yoon2014}, where the TR pre-filter is used in cascade configuration with a pre-equalizer, which is found by minimizing the ISI power through an semi-definite relaxation approximation. A multiuser TR equalization approach is found in \cite{viteri2016}, where the equalizer design is constrained to solutions that null the interference to other users. Reference \cite{yang2013} shows a TR waveform design that maximizes the sum rate in multiuser systems, and using a rate back-off strategy to reduce ISI. An equalized spatial multiplexing TR scheme for single-input multiple-output (SIMO) systems is presented in \cite{nguyen2010} for UWB.

However, previous approaches have not address the following aspects: \emph{i}) theoretical performance with respect to focusing capability or BER is not characterized, \emph{ii}) the behavior of previous solutions is not analyzed with respect to changes in propagation conditions, \emph{iii}) the required up-sampling for rate back-off in some solutions demands costly high-speed hardware and/or decreasing the transmission rate, and \emph{iv}) other solutions increase the receiver's computational complexity versus conventional TR by either using multiple receiving antennas or costly receiver equalizers. Thus, in the proposed ETR scheme, we use a discrete ZF pre-equalizer in cascade with a TR pre-filter at the transmitter in order to eliminate the ISI component in the received signal while preserving the spatial focusing of conventional TR beamforming. Our improvements with respect to previous works that have dealt with mitigating ISI in TR systems are:

\begin{itemize}

\item The proposed ETR technique adds computational complexity to the transmitter only, maintaining the simplicity of the conventional TR receiver. Such additional complexity compared to conventional TR is limited by using a single equalizer shared by all of the transmit antennas.

\item Unlike previous solutions, we derive theoretical performance bounds for both the probability of bit error and the beamforming capability of the proposed ETR technique. We also compare these bounds with those of conventional TR under different propagation conditions.

\end{itemize}

Our model is based on the assumption of a static (block fading) channel with perfect CSI at the transmitter. This assumption is particularly appropriate for indoor wireless communications, where the channel is slow-varying compared to the frame structure in upper layers. Therefore, we do not focus on the influence of channel estimation errors in our analysis. However, specific works regarding imperfect CSI at the transmitter in TR systems can be found in the literature \cite{alizadeh2012,alizadeh2013,jonietz2008,nfaqvi2011,tran2015}.
By means of numerical simulations, we validate the results and derived bounds herein under the assumed conditions. We also demonstrate that the proposed ETR technique outperforms conventional TR in terms of BER without a significant impact on the beamforming capability.

%%%%%%%%% Sumarize results %%%%%%%%%

%Using numerical simulations, we verify that our approximations are tight for the indoor scenarios considered here. We also demonstrate that a larger signal bandwidth increases the power at the receiver relative to other locations, but it increases ISI power as well. This applies to conventional wideband systems as well as to UWB systems. is the derivation of theoretical performance bounds for the BER and the spatial focusing parameter for ZFTR. We also compare the performance of conventional TR and ZFTR in typical indoor channels by means of numerical simulations, and verify that our bounds are satisfied.

The remainder of this paper is organized as follows. Section \ref{Model} describes the conventional TR and ETR system models. In Section \ref{Analysis}, we present the performance analysis of both techniques based on the power components of the received signal. We also define performance parameters and derive closed-form expressions for them. Section \ref{Results} presents numerical simulation results for the performance parameters and a comparison against the theoretical approximations. This is followed by concluding remarks in Section \ref{Conslusions}.

\section{System Model}
\label{Model}
\begin{figure}[!t]
\centering
\includegraphics[width=\columnwidth]{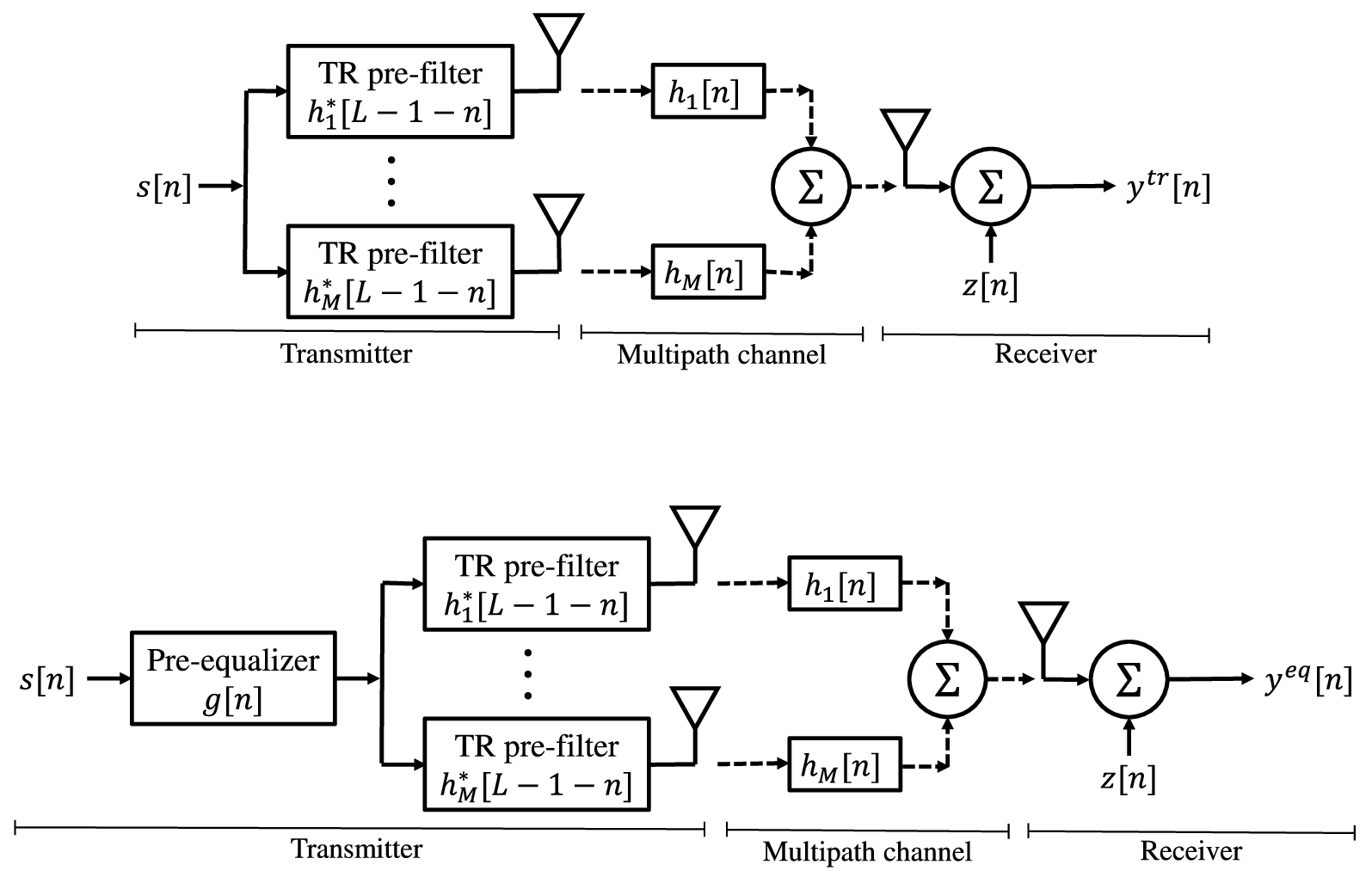}
\caption{Single-user MISO system model for conventional TR (up) and ETR (down). In conventional TR one pre-filter is used in each antenna. In ETR, an additional pre-equalizer is introduced to the transmitted in order to mitigate the ISI. Note that a minimum complexity receiver is used.}
\label{figmodel}
\end{figure}

In this section we introduce the discrete signal model for conventional TR and the proposed ETR. We also present the corresponding radio channel models that will be used in the next section to characterize the performance of those techniques. The general idea behind TR is to use the time-reversed CIR from every antenna to the receiver as a pre-filter for the transmitted signal. Such pre-filter acts as a beamformer in the spatial domain, focusing the RF power around the receiver. For the ETR case, we propose a TR pre-filter in cascade with a ZF pre-equalizer in order to mitigate the ISI of conventional TR. The system model for conventional TR and ETR is depicted in Fig. \ref{figmodel}.
\subsection{Conventional TR Signal Model}
\label{TRsignalmodel}
Consider a digital MISO baseband wireless communication system with $M$ transmit antennas and one single-antenna user. Let $s[n]$ be the complex transmitted signal representing arbitrarily modulated symbols, where $n \in \mathbb{Z}^+$ is the discrete time index. This signal is assumed to have unit power (i.e. $\mathbb{E}\left[ |s[n]|^2 \right]=1 \; \forall n$) regardless of the modulation. In conventional TR, the discrete time transmitted signal from the $i$-th antenna is
\begin{equation}
\label{eq_trtransmitted1}
x_i^{tr}[n] = \sqrt{\rho} \; s[n] \otimes \frac{h_i^*[L-1-n]}{\sqrt{P_h}} , \quad i = 1, \ldots , M, \nonumber
\end{equation}
where $\rho$ is the total average transmitted power, $\otimes$ denotes convolution; $h_i[n]$, $n = 0, \ldots, L-1$, is the complex CIR from the $i$-th transmit antenna to the receiver; and $P_h$ is a normalization factor introduced to ensure that the total transmitted power remains constant in every realization. This factor is defined as
\begin{equation}
\label{eq_channelpower}
P_h = \sum_{i=1}^{M} \sum_{l=0}^{L-1}|h_i[l]|^2.
\end{equation}
Then, $h_i^*[L-1-n]$ is the complex-conjugated time-reversed CIR applied as a pre-filter to the transmitted sequence. When perfect CSI is available at the transmitter and the channel is static, the received baseband signal is
\begin{IEEEeqnarray}{rCl}
\label{eq_trreceived}
y_{tr}[n] & = & \frac{1}{\sqrt{P_h}} \sum_{i=1}^{M} s[n] \otimes h_i^*[L-1-n] \otimes h_i[n] + z[n] \nonumber \\
	 	  & = & \sum_{i=1}^{M} s[n] \otimes  h_{tr,i} [n] + z[n] \nonumber
\end{IEEEeqnarray}
where $z[n]$ represents additive white Gaussian noise with variance $\sigma_z^2$, and we have defined the equivalent time-reversed CIR (TR-CIR) for the $i$-th antenna as
\begin{IEEEeqnarray}{rCl}
\label{eq_trcir}
h_{tr,i}[n] & = & \frac{1}{\sqrt{P_h}} h_i^*[L-1-n] \otimes h_i[n] \nonumber \\
			& = & \frac{1}{\sqrt{P_h}} \sum_{l=0}^{L-1} h_i[l] h_i^*[L-1-n+l] \nonumber \\
     		& & n = 0, \ldots , 2L-2. \nonumber
\end{IEEEeqnarray}
The effect of the TR filter is thus to replace the original CIR with the TR-CIR, whose properties will be analyzed hereinafter. Notice that we can rewrite the received signal in order to separate the desired symbol, the ISI, and the noise as
\begin{IEEEeqnarray}{rCl}
\label{eq_trreceived2}
y_{tr}[n] & = & \underbrace{ \sqrt{ \rho \sum_{i=1}^{M} \sum_{l=0}^{L-1} \left|h_i[l] \right|^2} \; s[n-L+1]}_{\text{desired symbol}} \nonumber \\
		  & & + \underbrace{\sqrt{\rho } \; \sum_{i=1}^{M} \sum_{\substack{ l=0 \\ l \neq L-1}}^{2L-2} h_{tr,i}[l] s[n-l] }_{\text{ISI}} + \underbrace{z[n]}_{\text{noise}}.
\end{IEEEeqnarray}
This separation in (\ref{eq_trreceived2}) can be interpreted in the following way. First, note that $h_{tr,i}[n]$ is a scaled autocorrelation function of $h_i[n]$, whose peak amplitude is
\begin{equation}
\label{eq_trcirpeak}
\max_n \, |h_{tr,i}[n]| = |h_{tr,i}[L-1]| = \frac{1}{\sqrt{P_h}} \sum_{l=0}^{L-1} | h_i[l] |^2. \nonumber
\end{equation}
Thus, the focusing time effected by TR occurs at sample $L-1$ in the TR-CIR. At that instant, the multipath components corresponding to the desired symbol add in phase, so its coefficient is real and positive. Moreover, the ISI components add incoherently. The net result is an increase in the desired signal power and a reduction in the ISI. Note that $h_{tr,i}[k]$ has $2L-1$ non-zero samples, so the ISI spans across $2L-2$ symbols{\footnote{{Note that, being an scaled autocorrelation function of $h_i[n]$, the equivalent time-reversed CIR has more samples than $h_i[n]$, but it has a lower delay spread (i.e. most of its energy is compressed in a number of samples less than $L$). In this paper, we take into account all of the non-zero samples of $h_{tr,i}[n]$ in order to fully characterize the residual ISI of TR beamforming.}}}.

%y_{tr}[n] & = & \underbrace{ \sqrt{ \rho \sum_{i=1}^{M} \sum_{l=0}^{L-1} \left|h_i[l] \right|^2} \; s[n-L+1]}_{\text{Desired Symbol}} \\
%&& + \underbrace{\sqrt{\rho} \; \sum_{i=1}^{M} \sum_{\substack{ l=0 \\ l \neq L-1}}^{2L-2} (h_i^*[L-1-l]\otimes h_i[l])s[n-l]}_{\text{ISI}} \nonumber \\
%&& + \underbrace{z[n]}_{\text{Noise}}

%%%%%%%%%%%%%%%%%%%%%%%%%%%%%%%%%%%%%%%%%%%%%%%%%%%%%%%%%%%%%%%%%%%%%%%%%%%%%%%%%%%%%%%%%%%%%%%%%
%%%%%%%%%%%%%%%%%%%%%%%%%%%%%%%%%%%%%%%%%%%%%%%%%%%%%%%%%%%%%%%%%%%%%%%%%%%%%%%%%%%%%%%%%%%%%%%%%
%%%%%%%%%%%%%%%%%%%%%%%%%%%%%%%%%%%%%%%%%%%%%%%%%%%%%%%%%%%%%%%%%%%%%%%%%%%%%%%%%%%%%%%%%%%%%%%%%

\subsection{Proposed Equalized TR Signal Model}
\label{ETRsignalmodel}
A main challenge in conventional TR beamforming is to mitigate the ISI component of the received signal. As seen in (\ref{eq_trreceived2}), and depending on the specific channel realization, the ISI can represent a significant percentage of the total received power, thus affecting detection. Typically this problem can be solved with equalization at the receiver, RAKE receivers or OFDM \cite{proakis2008}, but this would increase the low computational complexity enabled by TR. Thus, we propose an equalizer $g[n]$ of length $L_E$ (i.e. $n = 0, \ldots, L_E-1$) cascaded with the TR pre-filters, with the goal of minimizing the ISI power at the receiver \cite{fouda2012}, as shown in Fig. \ref{figmodel}. {A single equalizer is shared by all the transmit antennas in order to reduce the required computational complexity.} We refer to this approach as ETR. Following the model and notation in Section \ref{TRsignalmodel}, the ETR transmitted signal in the $i$-th antenna is
\begin{equation}
\label{eq_eqtransmitted}
x_i^{eq}[n] = \sqrt{\rho} \; s[n] \otimes \frac{h_i^*[L-1-n] \otimes g[n]}{\sqrt{P_g}}
\end{equation}
where the normalization factor $P_g$ is defined as
\begin{equation}
\label{eq_eqnormal}
P_g = \sum_{i=1}^{M} \sum_{n=0}^{L+L_E-2}\left|h_i^*[L-1-n] \otimes g[n]\right|^2. \nonumber
\end{equation}
Note that $P_g$ accounts for the number of antennas, so $\rho$ is not explicitly divided by $M$ in (\ref{eq_eqtransmitted}). When perfect CSI is available at the transmitter and the channel is static, the received signal in ETR is
\begin{equation}
\label{eq_eqreceived}
y_{eq}[n] = \sqrt{\frac{\rho}{P_g}} s[n] \otimes g[n] \otimes \sum_{i=1}^{M} h_{tr,i}[n] + z[n]. \nonumber
\end{equation}
We propose a ZF pre-equalizer design for $g[n]$ whose objective is to completely eliminate the ISI component in the received signal. Although ZF is vulnerable to noise when used at the receiver \cite{molisch2010}, this problem is not of concern at the transmitter. The ZF criterion for the equalizer design is
\begin{equation}
\label{eq_zfeqtime}
g_{zf}[n] \otimes \sum_{i=1}^{M} h_{tr,i}[n] = \delta[n-n_0],
\end{equation}
where $g[n]=g_{zf}[n]$ is the ZF equalizer solution, $\delta[n]$ is the unitary impulse function, and $n_0 \in [0,\ldots,2L+L_E-3]$ is an arbitrarily selected delay. Note that (\ref{eq_zfeqtime}) is a overdetermined system of linear equations with $L_E$ unknowns and $2L+L_E-2$ equations, which can be represented in matrix form as
%\begin{IEEEeqnarray}{rCl}
%\hspace{-10pt}
%\underbrace{\begin{bmatrix}
%\sum_{i=1}^{M} h_{tr,i}[0]		&	0								&	 \\
%\vdots							&	\sum_{i=1}^{M} h_{tr,i}[0]		&	\cdots \\
%\sum_{i=1}^{M} h_{tr,i}[2L-2]	&	\vdots							&	 \\
%0								&	\sum_{i=1}^{M} h_{tr,i}[2L-2]	&	 \\
%0								&	0								&	\ddots \\
%\vdots							&	\vdots							&	 \\
%\end{bmatrix}}_{\textstyle\mathbf{H}}
%\underbrace{
%\begin{bmatrix}
%g_{zf}[0]		\\
%\vdots		\\
%g_{zf}[L_E-1]	\\
%\end{bmatrix}}_{\textstyle\mathbf{g}_{zf}}=
%\underbrace{
%\begin{bmatrix}
%0	\\
%\vdots	\\
%0	\\
%1	\\
%0	\\
%\vdots	\\
%0	\\
%\end{bmatrix}}_{\boldsymbol{\textstyle\delta}_{n_0} }.
%\end{IEEEeqnarray}
\begin{IEEEeqnarray}{rCl}
\hspace{-10pt}
\underbrace{\begin{bmatrix}
\sum_{i=1}^{M} h_{tr,i}[0]		&	\\
\vdots							&	\ddots	\\
\sum_{i=1}^{M} h_{tr,i}[2L-2]	&	\ddots	\\
0								&	\ddots	\\
\vdots							&	\\
\end{bmatrix}}_{\textstyle\mathbf{H}}
\underbrace{
\begin{bmatrix}
g_{zf}[0]		\\
\vdots		\\
g_{zf}[L_E-1]	\\
\end{bmatrix}}_{\textstyle\mathbf{g}_{zf}}=
\underbrace{
\begin{bmatrix}
0	\\
\vdots	\\
0	\\
1	\\
0	\\
\vdots	\\
0	\\
\end{bmatrix}}_{\boldsymbol{\textstyle\delta}_{n_0}},\nonumber
\end{IEEEeqnarray}
where $\mathbf{H}\in \mathbb{C}^{(2L+L_E-2)\times L_E}$ is a banded Toepliz (convolution) matrix. Thus, the equalizer has only a least-squares solution $\mathbf{g}_{zf} = (\mathbf{H}^H\mathbf{H})^{-1}\mathbf{H}^H \boldsymbol{\delta}_{n_0}$, where (\ref{eq_zfeqtime}) is only satisfied when $L_E \rightarrow \infty$ \cite{hayes1996}. We now take the ZF criterion to the frequency domain in order to facilitate the analysis in Section \ref{Analysis}. Let $G_{zf}[k]$ and $H_i[k]$ denote the discrete Fourier transforms (DFT) of $g_{zf}[n]$ and $h_i[n]$, respectively, with $n, k = 0,\ldots,2L+L_E-3$ (zero padding is used in order to represent the linear convolution). After applying the DFT to (\ref{eq_zfeqtime}), the ZF equalizer in the frequency domain is
\begin{equation}
\label{eq_equalizerfreq}
G_{zf}[k] = \frac{e^{-j\frac{2\pi(n_0-L+1)}{2L+L_E-2}k}}{\sum_{i=1}^{M} \left| H_i[k] \right|^2}. 
\end{equation}
In the next section, we use the frequency domain representation given by (\ref{eq_equalizerfreq}) in order to obtain performance bounds for ETR. We also analyze the effect of equalizer's length $L_E$ over the ISI power. Using the ZF equalizer, the received signal is then
\begin{equation}
\label{eq_zfreceived}
y_{eq}[n] \approx \sqrt{\frac{\rho}{P_g}} \; s[n-n_0] + z[n],
\end{equation}
where the ISI term is neglected by assuming a sufficiently large $L_E$. This approximation is also analyzed in Section \ref{Results}.

%We can perform a matrix representation of the ZF condition as
%\begin{equation}
%\label{eq_zfeqtoeplitz}
%\mathbf{Hg} = \mathbf{\Delta_0}
%\end{equation}
%where $\mathbf{H} \in \mathbb{C}^{2L+L_E-2 \times L_E}$ is a Toeplitz matrix with first column vector equal to $\left(h[0],\ldots,h[2L-2], 0,\ldots,0\right)^T$, $h[n]=\sum_{i=1}^{M} h_{tr,i}[n]$, $\mathbf{g} \in \mathbb{C}^{L_E}$ is the vector representing the equalizer's coefficients, and $\mathbf{\Delta_0} = \left(0,\ldots,0,1,0,\ldots,0\right)$, with $1$ in the $n_0+1$ position. Note that (\ref{eq_zfeqtoeplitz}) is an overdetermined linear system of equations, whose least squares solution is given by
%\begin{equation}
%\label{eq_zfsolution}
%\mathbf{g}_{ls} = \left(\mathbf{H}^H \mathbf{H}\right)^{-1} \mathbf{H}^H \mathbf{\Delta_0},
%\end{equation}
%where the superscript $^H$ denotes complex conjugate transpose.

%%%%%%%%%%%%%%%%%%%%%%%%%%%%%%%%%%%%%%%%%%%%%%%%%%%%%%%%%%%%%%%%%%%%%%%%%%%%%%%%%%%%%%%%%%%%%%%%%
%%%%%%%%%%%%%%%%%%%%%%%%%%%%%%%%%%%%%%%%%%%%%%%%%%%%%%%%%%%%%%%%%%%%%%%%%%%%%%%%%%%%%%%%%%%%%%%%%
%%%%%%%%%%%%%%%%%%%%%%%%%%%%%%%%%%%%%%%%%%%%%%%%%%%%%%%%%%%%%%%%%%%%%%%%%%%%%%%%%%%%%%%%%%%%%%%%%

\begin{table*}[t]
\caption{Channel Model Parameters}%
\label{TableChannelModel}
\centering
\begin{tabular}{ccccccccc}
\hline \hline
\multicolumn{1}{c}{\multirow{2}{*}{\centering{Tap Separation ($T_s$) [ns]}}} & \multicolumn{2}{c}{1 cluster} & \multicolumn{6}{c}{2 clusters} \\
& $\sigma$ [ns] & $L$ & $\gamma$ & $\sigma_1$ [ns] & $\sigma_2$ [ns] & $L_1$ & $L_2$ & $L$ \\
 \hline
2.5 	& 8 & 33 & $0.4786$ & 8 & 14 & 8 & 17 &33\\
5 		& 8 & 17 & $0.4786$ & 8 & 14 & 4 & 9 & 17\\
10 		& 8 & 9  & $0.4786$ & 8 & 14 & 2 & 5 & 9\\
\hline \hline
\end{tabular}
\end{table*}

\subsection{Wideband Radio Channel Model}
\label{channelmodel}
As  mentioned above, TR benefits from rich scattering, so it can be conveniently applied for indoor wireless communications. We selected two statistical baseband channel models suitable for such scenarios to make the performance analysis. The first one is a simple single-cluster CIR model with exponential power decay in time. The second model is a more general case with two propagation clusters, each one of them with exponential power decay. Even though the first model is a particular case of the second, we consider it here separately in order to illustrate the derivation process and to facilitate interpretation of the results in Section \ref{Analysis}. {In addition, as demonstrated in Section \ref{Analysis}, the performance of TR is strongly dependent on propagation conditions, i.e. PDP and delay spread. Most of the results of current literature use only the single-cluster channel model for the analysis of TR techniques. However, by using a second PDP, we show that the analyses of TR performance are model-dependent and should not be generalized.}

For simplicity, we only take into account here the case where each CIR tap represents the contribution from several unresolvable multipath components with the same average amplitudes. Thus, diffuse scattering is assumed and both channel models have Rayleigh distributions. The common features of the two models are that the CIR $h_i[n]$ is modeled as a circular symmetric complex Gaussian random variable with zero mean $\forall i, n$. We assume that the transmit array elements have sufficient separation (e.g. irregular array). The system operates in a rich scattering environment, so $h_i[n]$ and $h_{i'}[n']$ are independent and uncorrelated if $i \neq i'$ or $n \neq n'$ (i.e. uncorrelated scattering). We also define the following constraint on the CIR total power:
\begin{equation}
\label{eq_channelnormalization}
\sum_{l=0}^{L-1} \mathbb{E}\left[ \left| h_i[l] \right|^2 \right] = \Gamma, \ \forall i,
\end{equation}
where $\Gamma \ll 1$ is a constant accounting for the channel induced propagation losses. This constraint implies that the channels between each transmit antenna and the receiver have the same average power. The variance of $h_i[n]$ is specified by the power delay profile (PDP) model, as follows:

\subsubsection{Model 1}
This is the standard reference PDP model for indoor wireless communications \cite{saleh1987}. The power in the CIR decreases exponentially in time with a single scattering cluster:
\begin{equation}
\label{eq_model1}
\mathbb{E} \left[ |h_i[n]|^2 \right] = \left\{
\begin{array}{ll}
A e^{-\frac{nT_s}{\sigma}} & \text{if } n=0,\ldots,L-1, \\
0 & \text{otherwise,}
\end{array} \right. \nonumber
\end{equation}
where $T_s$ is the sampling period or tap spacing, $\sigma$ is the delay spread parameter, and $A$ is selected to satisfy (\ref{eq_channelnormalization}).

\subsubsection{Model 2}
The PDP matches common indoor propagation models, such as the IEEE 802.11n/ac Channel B in \cite{erceg2004} and \cite{breit2009}. This is an exponential decay model with two scattering clusters. This is valid for indoor WLANs with operating frequencies around 2.4 GHz and 5 GHz, and bandwidths of up to 1.28 GHz:
\begin{equation}
\label{eq_model2}
\mathbb{E} \left[ |h_i[n]|^2 \right] = \left\{
\begin{array}{ll}
A e^{-\frac{nT_s}{\sigma_1}} & \text{if }  0 \leq n \leq L_1 - 1, \\
A e^{-\frac{nT_s}{\sigma_1}} + \gamma \, A \, e^{-\frac{(n-L_1)T_s}{\sigma_2}} & \text{if }  L_1 \leq n \leq L_2 - 1, \\
\gamma \, A  \, e^{-\frac{(n-L_1)T_s}{\sigma_2}} & \text{if } L_2 \leq n \leq L - 1,\\
0 & \text{otherwise,}
\end{array} \right. \nonumber
\end{equation}
where $\sigma_1$ and $\sigma_2$ are the delay spread parameters, $L_1$ is the starting sample for the second cluster, $L_2$ is the number of samples in the first cluster, $\gamma$ is the relative power of the second cluster, and $A$ is the normalization constant selected such that (\ref{eq_channelnormalization}) is satisfied.

Note that both models correspond to Rayleigh channels, with a duration of $L$ samples in the CIR. However, Model 2 has a higher delay spread due to the strong delayed power contribution from the second scattering cluster. Table \ref{TableChannelModel} shows the parameter values of each channel model under different CIR lengths, selected according to the standard \cite{erceg2004}, \cite{breit2009}. Parameters for Model 1 are the same as those for the first cluster in Model 2. These parameters are used for comparison purposes in Section \ref{Results}.

%\item In previous works \cite{nguyen2006,wang2011,han2012}, an statistical channel model with exponential power decay in time was used for the performance analysis of TR systems. Although this model is suitable for indoor scenarios, we also use a more general model that allows for the comparison under different propagation conditions (e.g. richer scattering). Thus, our derivations use both channel models with the first one being a particular case of the second. 

\section{Performance Analysis of Conventional TR and ETR}
\label{Analysis}

We now characterize the performance of conventional TR and the proposed ETR technique with respect to the probability of bit error and the spatial focusing capability. In conventional TR, as stated Section \ref{TRsignalmodel}, the received signal (\ref{eq_trreceived2}) has three components: desired symbol, ISI, and noise. Individual components in the ISI sum have a complex double gaussian distribution \cite{odonoughue2012}, and they are dependent random variables. Thus, the ISI sum does not meet the assumptions of the conventional central limit theorem, and its distribution does not necessarily converges to a Gaussian distribution when the number of terms goes to infinity \cite{hilhorst2009}. Nevertheless, an approximation to the probability of bit error in conventional TR systems assuming that ISI is Gaussian has been found to be sufficiently close to the numerical results in Section \ref{Results}. For BPSK and QPSK  modulations, this approximation is
%Although individual components of ISI are not Gaussian, they are independent and meet the Lindberg condition\footnote{\color{red}The central limit theorem requires the terms in the sum to be iid. However, by meeting the Lindeberg condition, those terms are not required to be identically distributed.} \cite[p.213]{stark1994}, so the ISI sum is approximately Gaussian due to the Central Limit Theorem. This approximation improves by increasing L and/or M. Therefore, the probability of bit error in conventional TR systems for BPSK and QPSK can be written as
\begin{IEEEeqnarray}{lCl}
\label{eq_pe1}
P_{e,BPSK}^{tr}  \approx  Q \left( \sqrt{\frac{2P_{S}}{P_{ISI}+\sigma_z^2}} \right)\quad \text{and} \nonumber \\
  P_{e,QPSK}^{tr} \approx Q \left( \sqrt{\frac{P_{S}}{P_{ISI}+\sigma_z^2}} \right),
\end{IEEEeqnarray}
respectively, where  $Q(\cdot)$ is the complementary cumulative distribution function of a standard Gaussian random variable, $P_S$ is the desired signal power, $P_{ISI}$ is the inter-symbol interference power, and $\sigma_z^2$ is the noise power. Note that in conventional TR the performance is limited by ISI at high SNR. {In the case of ETR, we assume that we can neglect the ISI term in the received signal due to equalization, which is true for a sufficiently large $L_E$ (as analyzed next). Thus, a lower bound on the probability of bit error in BPSK and QPSK \cite{proakis2008} using ETR are, respectively,
\begin{IEEEeqnarray}{rCl}
\label{eq_pe3}
& P_{e,BPSK}^{eq}  \geq  Q \left( \sqrt{2\frac{P_{eq}}{\sigma_z^2}} \right) \quad \text{and} \quad P_{e,QPSK}^{eq}  \geq  Q \left( \sqrt{\frac{P_{eq}}{\sigma_z^2}} \right),& \nonumber
\end{IEEEeqnarray}
where $P_{eq}$ is the received signal power in (\ref{eq_zfreceived}). Similar expressions for other modulations can be found in \cite{proakis2008}.} In this section, we derive the expressions for the power of each of those components in terms of the channel PDP, which are necessary for the performance characterization of TR and ETR. These expressions have not been compared previously across different channel models, so they constitute one of the contributions of this paper. We also study the influence of the equalizer's length over its ISI suppression capability. In addition, we define parameters to measure the TR space-time focusing performance, and then present closed-form approximations for them using the indoor channel models introduced above.

\subsection{Desired Signal Power}

\subsubsection{Conventional TR}
The desired signal power in (\ref{eq_trreceived2}) is
\begin{IEEEeqnarray}{rCl}
\label{eq_signalpower}
P_{S} & = & \mathbb{E} \left[  \rho \sum_{i=1}^M \sum_{l=0}^{L-1} \left| h_i[l] \right|^2   \right] = \rho \, M \, \Gamma.
\end{IEEEeqnarray}
which can be obtained from the channel power constraint (\ref{eq_channelnormalization}). Note that this signal power is independent of the channel model and is directly proportional to the number of antennas.

\subsubsection{ETR}
According to (\ref{eq_zfreceived}), the received signal power is
\begin{equation}
\label{eq_zfreceivedpower}
P_{eq} = \rho \, \mathbb{E} \left[\frac{1}{P_g}\right]. \nonumber
\end{equation}
As shown in Appendix \ref{AppA}, an upper bound on the received power (which causes a lower bound in the probability of bit error) is
\begin{equation}
\label{eq_zfreceivedpowerbound}
P_{eq} \leq \rho \, M \, \Gamma.
\end{equation}
Thus, the received power in ETR is at best the desired power in conventional TR and a reduction in the beamforming capability is expected. We analyze this issue later. However, the probability of bit error is lower in ETR due to the elimination of the ISI. We verify this bound numerically in Section \ref{Results}.

\subsection{Intersymbol Interference Power in Conventional TR}
The ISI power $P_{ISI}$, is derived here from (\ref{eq_trreceived2}) as the sum of the power in the TR-CIR at instants other than the focusing time (i.e., $l\in\{0,\ldots,2L-2\}$, $l \neq L-1$):
\begin{IEEEeqnarray}{rCl}
\label{eq_ISIpower}
P_{ISI} & = & \rho \, \mathbb{E} \left[ \left| \sum_{i=1}^{M} \sum_{\substack{ l=0 \\ l \neq L-1}}^{2L-2} h_{tr,i}[l] \right|^2 \right] \nonumber \\
        & = & \rho \, \mathbb{E} \left[ \left| \sum_{i=1}^{M} \sum_{\substack{ l=0 \\ l \neq L-1}}^{2L-2} \frac{h_i^*[L-1-l] \otimes h_i[l]}{\sqrt{P_h}} \right|^2 \right]. 
\end{IEEEeqnarray}
Note that $P_h$ is a random variable that depends on the CIR, as given by (\ref{eq_channelpower}), so the calculation of (\ref{eq_ISIpower}) is not straightforward. As shown in \ref{AppB}, we use an expansion for the expectation of the ratio of correlated random variables \cite{rice2008} \cite{rice2009} in order to derive the following approximation for this equation:
\begin{IEEEeqnarray}{lCl}
\label{eq_ISIpower1}
\hat{P}_{ISI} = \nonumber \\
\frac{\rho}{M \Gamma} \sum_{\substack{ l=0 \\ l \neq L-1}}^{2L-2} \sum_{i=1}^{M} \sum_{\substack{ n=0 \\ n \leq l \\ n \geq l-L+1 }}^{L-1} \mathbb{E} \left[ \left| h_i[n] \right|^2 \right]  \mathbb{E} \left[ \left| h_i[L-1-l+n] \right|^2 \right]. \nonumber \\ &&
\end{IEEEeqnarray}
{The error of this approximation reduces with a small number of antennas, larger delay spreads, and/or larger bandwidths (see Appendix \ref{AppB}).} Notice that the received ISI power depends on the PDP model. Therefore, we evaluate (\ref{eq_ISIpower1}) using the channel models described in Section \ref{channelmodel}. From now on, let the superscripts $^{(1)}$ and $^{(2)}$ denote variables calculated using Model 1 and Model 2, respectively, and the symbol $\hat{}$ denote the corresponding variable approximation. Then, the results for $P_{ISI}$ are as shown in (\ref{eq_ISIpower21}) and (\ref{eq_ISIpower22}), respectively.
\begin{IEEEeqnarray}{rCl}
\label{eq_ISIpower21}	
\hat{P}_{ISI}^{(1)} = \rho \, \Gamma \left( \frac{1 - e^{-\frac{T_s}{\sigma}}}{1 - e^{-\frac{LT_s}{\sigma}}} \right)^2 \sum_{\substack{ l=0 \\ l \neq L-1}}^{2L-2} \sum_{\substack{ n=0 \\ n \leq l \\ n \geq l-L+1 }}^{L-1} e^{-\frac{(L-1-l+2n)T_s}{\sigma}}, \nonumber \\
\end{IEEEeqnarray}
\begin{figure*}[t]
\normalsize % ensure that we have normalsize text
\setcounter{MYtempeqncnt}{\value{equation}} % Store the current equation number.
% Set the equation number to one less than the one
% desired for the first equation here.
% The value here will have to changed if equations
% are added or removed prior to the place these
% equations are referenced in the main text.
\setcounter{equation}{13}
\begin{IEEEeqnarray}{rCl}
\label{eq_ISIpower22}	
\hat{P}_{ISI}^{(2)}& = & \rho \, \Gamma  \, \frac{\sum\limits_{\substack{ l=0 \\ l \neq L-1}}^{2L-2} \left( \sum\limits_{\substack{ n=0, n \leq l \\ n \geq l-L+1 }}^{L_2-1} e^{-\frac{nT_s}{\sigma_1}} C[l,n] + \gamma \, \sum\limits_{\substack{ n=L_1, n \leq l \\ n \geq l-L+1 }}^{L-1} e^{-\frac{(n-L_1)T_s}{\sigma_2}} C[l,n] \right)}{\left( \sum\limits_{n=0}^{L_2-1} e^{-\frac{nT_s}{\sigma_1}} + \gamma \sum\limits_{n=L_1}^{L-1} e^{-\frac{(n-L_1)T_s}{\sigma_2}}\right)^2},\nonumber \\
\end{IEEEeqnarray}
\begin{equation}
\label{eq_ISI3}
C[l,n] = \left\{
\begin{array}{ll}
e^{-\frac{(L-1-l+n)T_s}{\sigma_1}}& \text{if } l-L+1 \leq n \leq l-L+L_1, \\
e^{-\frac{(L-1-l+n)T_s}{\sigma_1}} + \gamma e^{-\frac{(L-1-l+n-L_1)T_s}{\sigma_2}} & \text{if } l-L+L_1+1 \leq n \leq l-L+L_2, \\
e^{-\frac{(L-1-l+n-L_1)T_s}{\sigma_2}}  & \text{if } l-L+L_2+1 \leq n \leq l, \\
0 & \text{otherwise}.\\
\end{array} \right. \nonumber
\end{equation}
\hrulefill
\setcounter{equation}{15}
\begin{IEEEeqnarray}{rCl}
\label{eq_UsablePower2}	
\hat{U}^{(2)}& = & M \frac{\left( \sum\limits_{n=0}^{L_2-1} e^{-\frac{nT_s}{\sigma_1}} + \gamma \sum\limits_{n=L_1}^{L-1} e^{-\frac{(n-L_1)T_s}{\sigma_2}}\right)^2}{\sum\limits_{\substack{ l=0 \\ l \neq L-1}}^{2L-2} \left( \sum\limits_{\substack{ n=0, n \leq l \\ n \geq l-L+1 }}^{L_2-1} e^{-\frac{nT_s}{\sigma_1}} C[l,n] + \gamma \, \sum\limits_{\substack{ n=L_1, n \leq l \\ n \geq l-L+1 }}^{L-1} e^{-\frac{(n-L_1)T_s}{\sigma_2}} C[l,n] \right)}.
\end{IEEEeqnarray}
\hrulefill
\end{figure*}
\setcounter{equation}{14}
There are two interesting remarks about the power components in conventional TR that we found through the proposed approximation. First, the ISI power does not depend on the number of antennas, but the desired signal power is directly proportional to it. Hence, from the probability of error (\ref{eq_pe1}), an increase in $M$ would increase the ratio between $P_S$ and $P_{ISI}$ and, consequently, it would improve the BER at high SNR. This phenomena could be harnessed in the context of massive MIMO systems \cite{viteri2015}. {Second, there are three parameters that can affect the ISI power: the tap separation $T_s$ (or, equivalently, the bandwidth), the channel delay spread $\sigma$, and the CIR duration $L$. Thus, ISI power is strongly dependent on the propagation environment. In order to obtain a better insight on the impact of these three parameters on the BER performance of TR beamforming, we next define the usable power ratio relating desired signal power and ISI power.}

\subsection{Usable Power and Time Compression in Conventional TR}
The usable power ratio is a parameter that will help to compare different scenarios (characterized by their channel models) through a single metric. From the received signal in (\ref{eq_trreceived2}), we know that the total received power with conventional TR is $P_{R} = P_S + P_{ISI}$. Note that, according to the probability of error (\ref{eq_pe1}), conventional TR performance is limited by the ratio between $P_S$ and $P_{ISI}$ in the high SNR regime. Thus, the usable power ratio is defined as $U \triangleq P_S/P_{ISI}$, which measures the fraction of the received power that can be effectively used at the detector and determines a lower bound to (\ref{eq_pe1}). Using the expressions for Model 1 and Model 2, the usable power ratio approximations are given by (\ref{eq_UsablePower1}) and (\ref{eq_UsablePower2}), respectively.
\begin{IEEEeqnarray}{rCl}
\label{eq_UsablePower1}
\hspace*{-20pt} \hat{U}^{(1)} & = & \frac{M\left(1-e^{-\frac{LT_s}{\sigma}}\right)^2}{\left(1-e^{-\frac{T_s}{\sigma}}\right)^2 \sum\limits_{\substack{ l=0 \\ l \neq L-1}}^{2L-2} \sum\limits_{\substack{ n=0 \\ n \leq l \\ n \geq l-L+1 }}^{L-1} e^{-\frac{(L-1-l+2n)T_s}{\sigma}}},
\end{IEEEeqnarray}
This particular parameter has no relevance for ETR, since we assume the equalizer completely eliminates ISI. { We analyze numerically the impact of propagation conditions (namely, parameters $T_s$, $\sigma$, and $L$) over $U$ in Section \ref{Results}.}

\subsection{Interference Mitigation and Spatial Focusing}

The spatial focusing capability of conventional TR has important interference mitigation applications in wireless communications. In this subsection we analyze the signal power at points in the space different than the receiver's location by considering an unintended receiver with uncorrelated CIR. { Physically, in the frequencies where the employed channel models are valid, uncorrelated CIRs are obtained with just a few wavelengths of separation (e.g. see \cite{kafle2008}). We use this analysis to determine the power ratio between the targeted receiver's and nearby locations as a measure of the spatial focusing, and compare conventional TR with our proposed ETR technique.

Consider an unintended receiver with CIR denoted by $h_{u,i}[n]$ from the $i$-th transmit antenna, where $h_{u,i}[n]$ and $h_p[l]$ are identically distributed and uncorrelated for all $i$, $p$, $n$, and $l$. More specifically, $h_{u,i}[n]$ has the same power delay profiles and power constraints described in Section \ref{Model} for $h_i[n]$. In conventional TR, the signal at the unintended receiver is given by
\setcounter{equation}{16}
\begin{IEEEeqnarray}{rCl}
\label{eq_ureceiver}
y_u^{tr}[n] & = & \sqrt{\rho} \sum_{i=1}^{M} s[n] \otimes \frac{h_i^* [L-1-n]}{\sqrt{P_h}} \otimes h_{u,i}[n] + z[n]. \nonumber
\end{IEEEeqnarray}
The desired signal power captured by the unintended receiver is equal to the power of the sample at instant $L-1$ in its equivalent TR-CIR. Then, we define that interference power as
\begin{IEEEeqnarray}{rCl}
\label{eq_pint}
P_{int}^{tr} & = & \mathbb{E} \left[ \left| \sum_{i=1}^{M} \frac{h_i^* [L-1-n]}{\sqrt{P_h}} \otimes h_{u,i}[n] \right|^2 \right]_{n=L-1}. \nonumber
\end{IEEEeqnarray}
Using the same procedure that we used in the derivation of $P_{ISI}$, which can be found in Appendix \ref{AppB}, the interference power becomes
\begin{IEEEeqnarray}{rCl}
\label{eq_pint2}
\hat{P}_{int}^{tr} & = & \frac{\rho}{\Gamma} \sum_{l=0}^{L-1} \mathbb{E} \left[ \left| h_{u,i}[l]\right|^2 \right] \mathbb{E} \left[ \left| h_i [l] \right|^2 \right]. \nonumber
\end{IEEEeqnarray}
Again, this expression depends on the user PDP and the unintended receiver PDP, which are assumed to be identical. Thus, using the defined models, we get the results in (\ref{eq_pint31}) and (\ref{eq_pint32}).
\begin{IEEEeqnarray}{rCl}
\label{eq_pint31}
\hat{P}_{int}^{tr (1)} & = & \rho \, \Gamma \frac{\left( 1 + e^{-\frac{LT_s}{\sigma}} \right)\left( 1 - e^{-\frac{T_s}{\sigma}} \right)}{\left( 1 + e^{-\frac{T_s}{\sigma}} \right)\left( 1 - e^{-\frac{LT_s}{\sigma}} \right)},
\end{IEEEeqnarray}
\begin{figure*}[t]
\normalsize % ensure that we have normalsize text
\setcounter{equation}{17}
\begin{IEEEeqnarray}{rCl}
\label{eq_pint32}
\hat{P}_{int}^{tr(2)} & = & \rho \, \Gamma \, \frac{ \left( \sum\limits_{n=0}^{L_2-1} e^{-\frac{2nT_s}{\sigma_1}} + \gamma^2 \sum\limits_{n=L_1}^{L-1} e^{-\frac{2(n-L_1)T_s}{\sigma_2}}  + 2 \gamma \sum\limits_{n=L_1}^{L_2-1} e^{-\frac{nT_S}{\sigma_1}} e^{-\frac{(n-L_1)T_S}{\sigma_2}} \right)}{\left( \sum\limits_{n=0}^{L_2-1} e^{-\frac{nT_s}{\sigma_1}} + \gamma \sum\limits_{n=L_1}^{L-1} e^{-\frac{(n-L_1)T_s}{\sigma_2}}\right)^2}.
\end{IEEEeqnarray}
\hrulefill
\end{figure*}
In the proposed ETR technique, the signal at an unintended receiver is
\begin{IEEEeqnarray}{rCl}
\label{eq_intreceived}
y_{u}^{eq}[n] & = & \sqrt{\rho} \, s[n] \otimes \frac{g[n]}{\sqrt{P_g}} \otimes \sum_{i=1}^{M_T} h_i^* [n] \otimes h_{u,i}[n] +z[n]. \nonumber \\
\end{IEEEeqnarray}
In this case, the equalizer does not match the CIR to the unintended receiver, so the signal has a desired signal component and an ISI component due to imperfect equalization. This total received power can be approximated as (see Appendix \ref{AppC})
\begin{equation}
\label{eq_intpower}
\hat{P}_{int}^{eq} = \rho \, \Gamma.
\end{equation}
Note that, for the proposed ETR, both the received power and the interference power are independent of the channel model, as long as the power constraint (\ref{eq_channelnormalization}) is satisfied. We define the \emph{effective spatial focusing} parameter as the ratio between the \emph{usable} power at the receiver and the usable power at the unintended receiver (without considering the ISI in the signal). This parameter has been used previously in related literature, e.g. \cite{nguyen2006}. Then, for conventional TR and ETR this parameter is, respectively,
\begin{equation}
\label{eq_usablefucusing}
\eta_{tr} \triangleq \frac{P_S}{P_{int}^{tr}} \quad \text{and} \quad \eta_{eq} \triangleq \frac{P_{eq}}{P_{int}^{eq}}, \nonumber
\end{equation}
and measures the ability of the beamformer to focus the signal power on a specific point in space, {i.e. the power that can be used effectively at the detector.} In the case of conventional TR, we use the expressions (\ref{eq_signalpower}),  (\ref{eq_pint31}), and (\ref{eq_pint32}) to obtain the closed-form approximations to $\eta_{tr}$ in (\ref{eq_focusingtr1}) and (\ref{eq_focusingtr2}).
\begin{IEEEeqnarray}{rCl}
\label{eq_focusingtr1}
\hat{\eta}_{tr}^{(1)} & = & M \frac{\left( 1 + e^{-\frac{T_s}{\sigma}} \right)\left( 1 - e^{-\frac{LT_s}{\sigma}} \right)}{\left( 1 + e^{-\frac{LT_s}{\sigma}} \right)\left( 1 - e^{-\frac{T_s}{\sigma}} \right)},
\end{IEEEeqnarray}

\begin{figure*}[t]
\normalsize % ensure that we have normalsize text
\begin{IEEEeqnarray}{rCl}
\label{eq_focusingtr2}
\hat{\eta}_{tr}^{(2)} & = & M \frac{\left( \sum\limits_{n=0}^{L_2-1} e^{-\frac{nT_s}{\sigma_1}} + \gamma \sum\limits_{n=L_1}^{L-1} e^{-\frac{(n-L_1)T_s}{\sigma_2}}\right)^2}{ \left( \sum\limits_{n=0}^{L_2-1} e^{-\frac{2nT_s}{\sigma_1}} + \gamma^2 \sum\limits_{n=L_1}^{L-1} e^{-\frac{2(n-L_1)T_s}{\sigma_2}}  + 2 \gamma \sum\limits_{n=L_1}^{L_2-1} e^{-\frac{nT_S}{\sigma_1}} e^{-\frac{(n-L_1)T_S}{\sigma_2}} \right)}.
\end{IEEEeqnarray}
\hrulefill
\end{figure*}

{It is clear that the spatial focusing in TR increases with the number of antennas in a similar way as in a conventional phased array. Nevertheless, TR allows a 3D focusing of the signal using the information in the CIR, instead of the 2D beam-steering performed by phased arrays, i.e. TR can achieve full array gain in multipath environments. A numerical analysis of the behavior of $\eta_{tr}$ is given in Section \ref{Results} with respect to the channel model parameters. In the case of ETR, from (\ref{eq_zfreceivedpowerbound}) and (\ref{eq_intpower}), an upper bound on the spatial focusing parameter $\eta_{eq}$ is around $M$.}

We also define an alternate measure of spatial focusing that we call \emph{apparent power focusing}. This measures the total spatial focusing of the signal in conventional TR, including the presence of ISI. The definition is
\begin{equation}
\label{eq_totalfucusing}
\eta_{tr}' \triangleq \frac{P_S + P_{ISI}}{P_{int} + P_{ISI}}, \nonumber
\end{equation}
where the ISI power is the same at the unintended receiver, due to the fact that $h_{u,i}[n]$ and $h_{i}[n]$ have the same PDP. {In previous works, the difference between the \emph{effective power focusing} and the \emph{apparent power focusing} has not been clearly defined. Thus, we introduce this parameter in order to make a distinction between the total power present in the focusing point (which includes desired signal power and ISI), and the power that can be actually used at the detector (only the desired signal power).}

A detailed analysis of the parameters calculated in this section is provided next.

%P_{ISI}& \approx & \rho \left( \frac{1-e^{-\frac{T_s}{\sigma}}}{1-e^{-\frac{LT_s}{\sigma}}} \right)^2 \sum_{\substack{ k=0 \\ k \neq L-1}}^{2L-2} \sum_{\substack{ l=0 \\ l \leq k \\ l \geq k-L+1 }}^{L-1}  e^{-\frac{(L-1-k+2l)T_S}{\sigma}}

%\begin{figure*}[!t]
%\centering
%\subfloat[]{\includegraphics[width=3.5in]{Pe40_1.eps}
%\label{fig_401}}
%\subfloat[]{\includegraphics[width=3.5in]{Pe40_2.eps}
%\label{fig_402}}
%\newline
%\subfloat[]{\includegraphics[width=3.5in]{Pe80_1.eps}
%\label{fig_801}}
%\subfloat[]{\includegraphics[width=3.5in]{Pe80_2.eps}
%\label{fig_802}}
%\newline
%\subfloat[]{\includegraphics[width=3.5in]{Pe160_1.eps}
%\label{fig_1601}}
%\subfloat[]{\includegraphics[width=3.5in]{Pe160_2.eps}
%\label{fig_1602}}
%\newline
%\caption{Results for (a) 40 MHz bandwidth - Model 1, (b) 40 MHz bandwidth - Model 2, (c) 80 MHz bandwidth - Model 1, (d) 80 MHz bandwidth - Model 2, (e) 160 MHz bandwidth - Model 1, (f) 160 MHz bandwidth - Model 2. The difference between the theoretical $P_e$ approximation and the simulated BER results decreases when either bandwidth or delay spread increases. It is also observed that the bit error performance degrades for larger bandwidths due to the decrease in the usable power.}
%\label{fig_sim}
%\end{figure*}

\section{Numerical Results and Discussion}
\label{Results}

In this section, we illustrate the time compression property of TR and ETR by analyzing their equivalent CIRs. Then, we present numerical results for the performance parameters defined in Section \ref{Analysis}.

\subsection{Time Compression and Pre-Equalization}
Fig. \ref{fig_cir} shows the time compression property of conventional TR and ETR. The original CIRs (one per transmit antenna) have power contributions from all the multipath components at different times. TR beamforming focuses the all those contributions in a single sampling instant, but there is a significant residual ISI power. ETR mitigates the ISI at the cost of a reduced focusing on the desired sampling instant, so the equivalent CIR approaches a delta function. Moreover, the ISI power is diminishingly small as $L_E \rightarrow \infty$. Fig. \ref{fig_le} shows the behavior of desired signal power and ISI power as a function of equalizer length $L_E$. These results were obtained by averaging those powers over 1000 channel realization using Model 2 with $M=4$, $L=33$, and $T_s=2.5$ ns. {Both signal power and ISI power decay exponentially as $L_E$ increases, until no significant variation is observed. This occurs when $L_E \approx L$, which corresponds to a ratio of approximately 30 dB between signal power and ISI power in the worst case.  These results indicate that near-cancellation of ISI is achieved with a finite equalizer's length. If the number of antennas is increased, the required equalizer's length decreases proportionally, as can be concluded from the usable power parameter $U$ definition in Section \ref{Analysis}. Therefore, we set $L_E = L$ for the following simulations in this section, noting that such equalizer's length allows the system to be noise limited rather than ISI limited.}
\begin{figure}[!t]
\centering
\subfloat[]{\hspace{-15pt}\includegraphics[width=0.33\columnwidth]{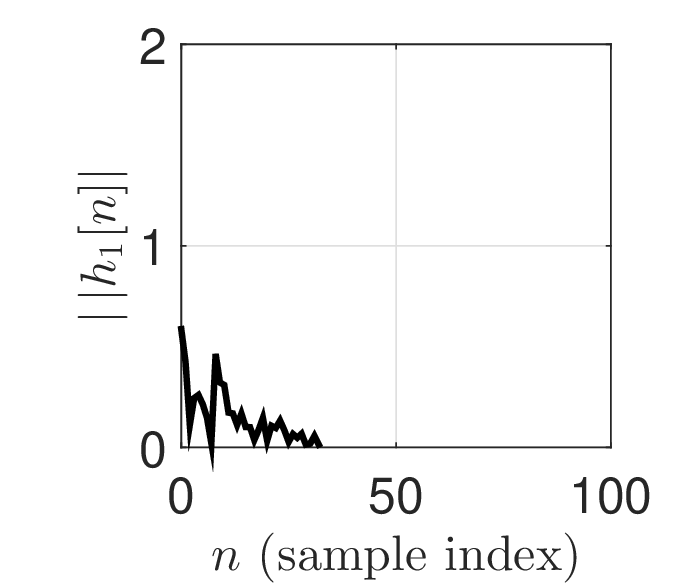}}
\subfloat[]{\includegraphics[width=0.33\columnwidth]{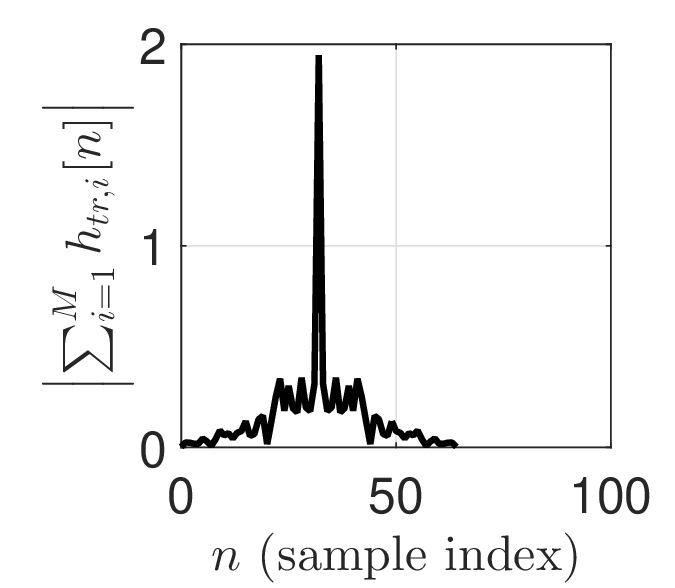}}
\subfloat[]{\includegraphics[width=0.33\columnwidth]{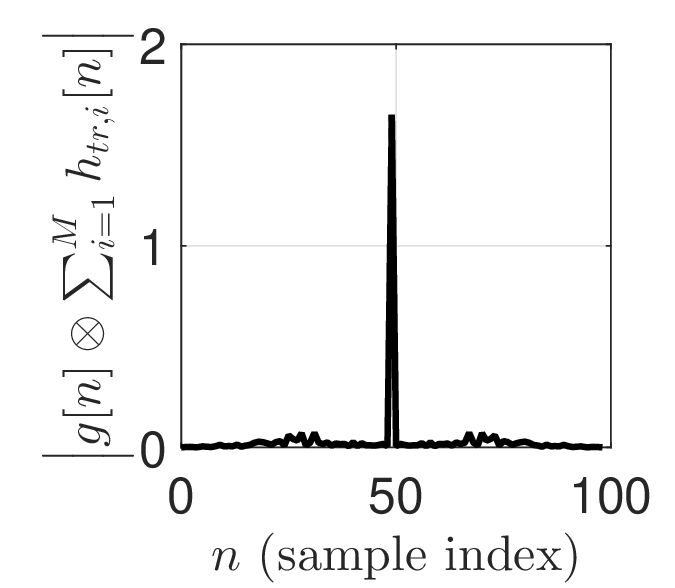}}
\caption{(a) one CIR realization for antenna 1 generated according to Model 2 with $L=33$. (b) equivalent TR-CIR obtained with conventional TR, i.e. as observed by the receiver; note the time focusing capability at the 32-th sample. (c) ETR equivalent CIR: a ZF pre-equalizer with length $L_E=33$ is cascaded with the TR pre-filters. ISI is greatly reduced with this approach at the cost of a reduced focusing. The equivalent CIR approaches a delta function. Results with $M=4$ antennas.}
\label{fig_cir}
\end{figure}
\begin{figure}[t]
%\centering
\subfloat[]{\hspace*{-5pt}\includegraphics[width=0.53\columnwidth]{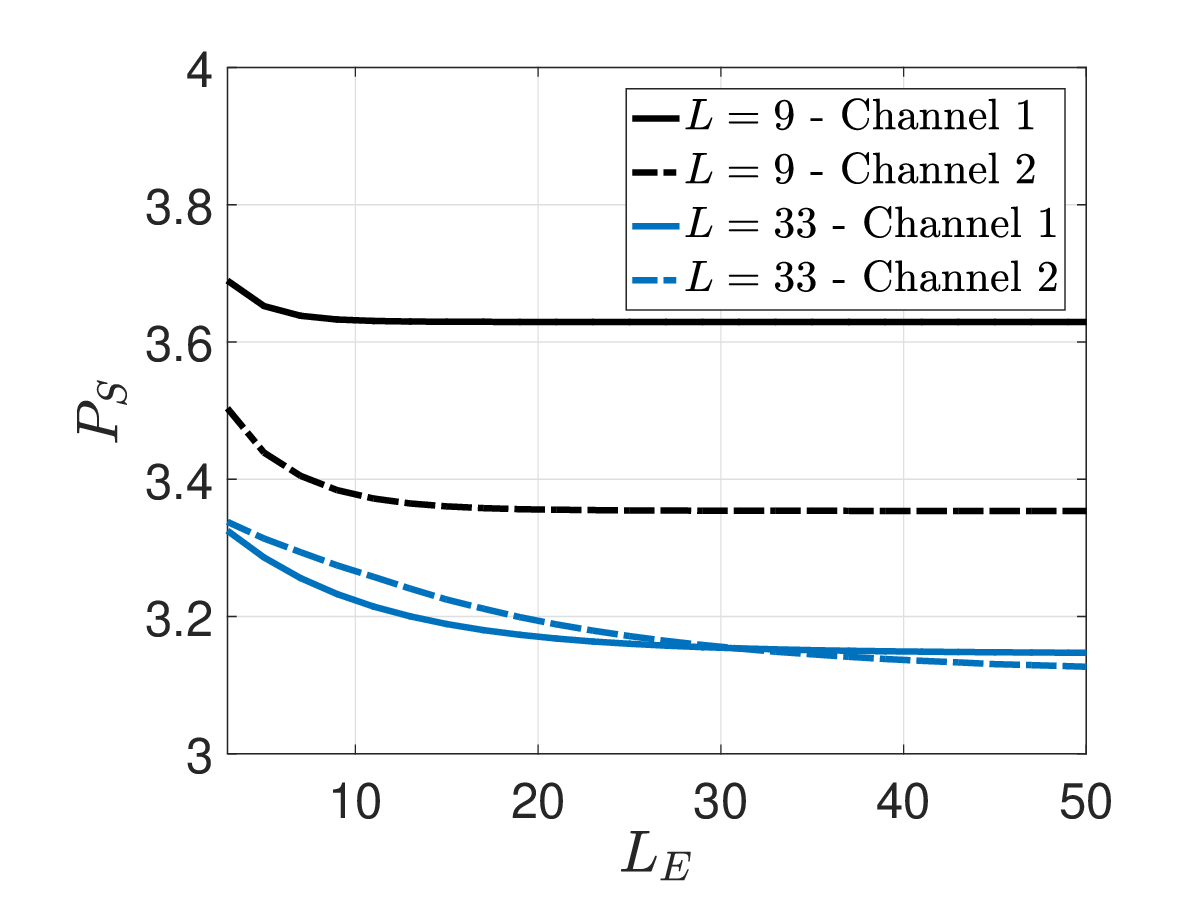}}
\subfloat[]{\hspace*{-10pt}\includegraphics[width=0.53\columnwidth]{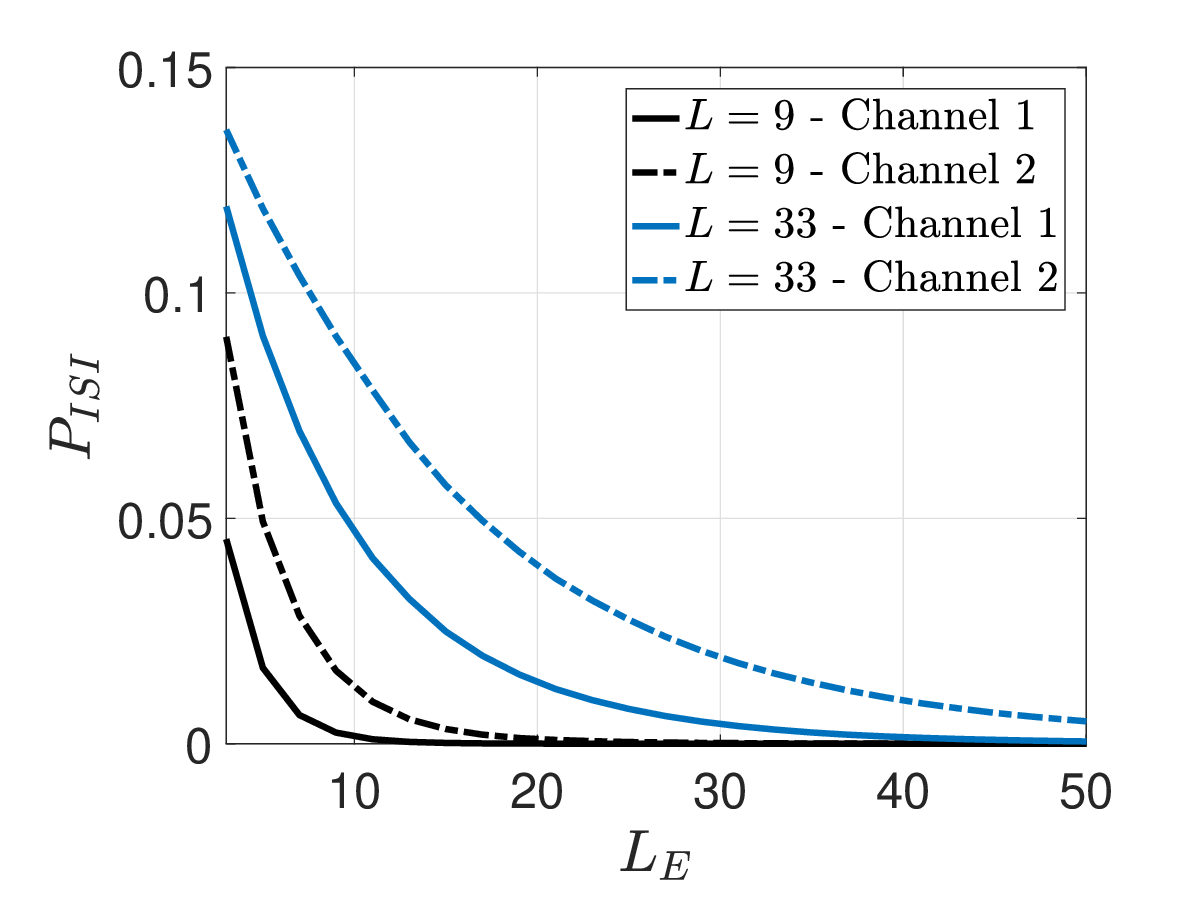}}
\caption{(a) Desired signal power, and  (b) ISI power as a function of equalizer's length $L_E$. Note that ISI power is larger when CIR length and/or delay spread increases (i.e. Model 2). Both powers decay by increasing $L_E$, and nearly perfect ISI suppression can be achieved at the cost of a marginal decrease in desired signal power. Results with $M=4$.}
\label{fig_le}
\end{figure}

\subsection{Beamforming Performance Parameters}

\begin{figure}[t]
\centering
\subfloat[]{\hspace*{-10pt}\includegraphics[width=1.11\columnwidth]{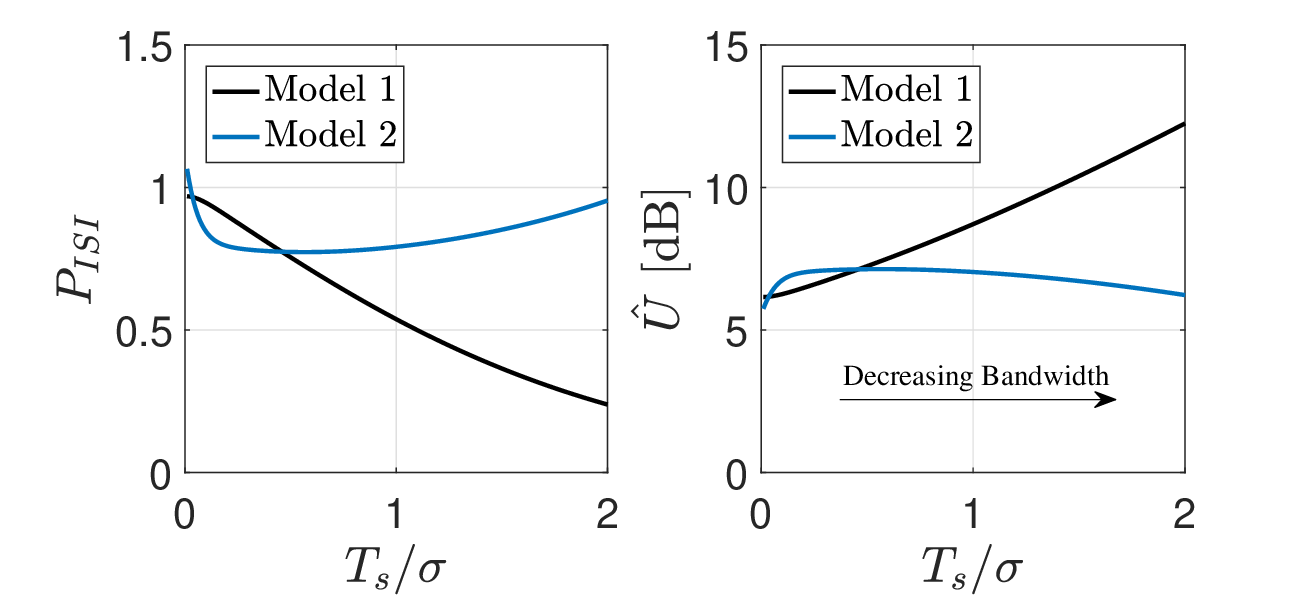}
\label{fig_U}}
\newline
\centering
\subfloat[]{\hspace*{-10pt}\includegraphics[width=1.11\columnwidth]{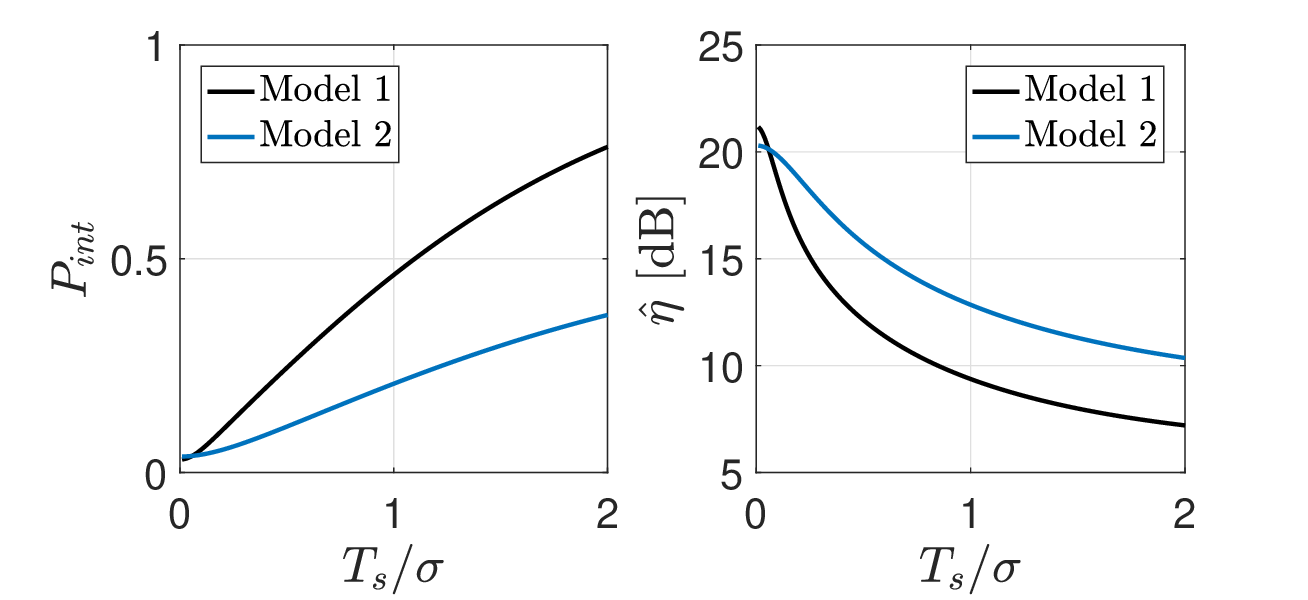}
\label{fig_Eta}}
\newline
\centering
\subfloat[]{\includegraphics[width=0.55\columnwidth]{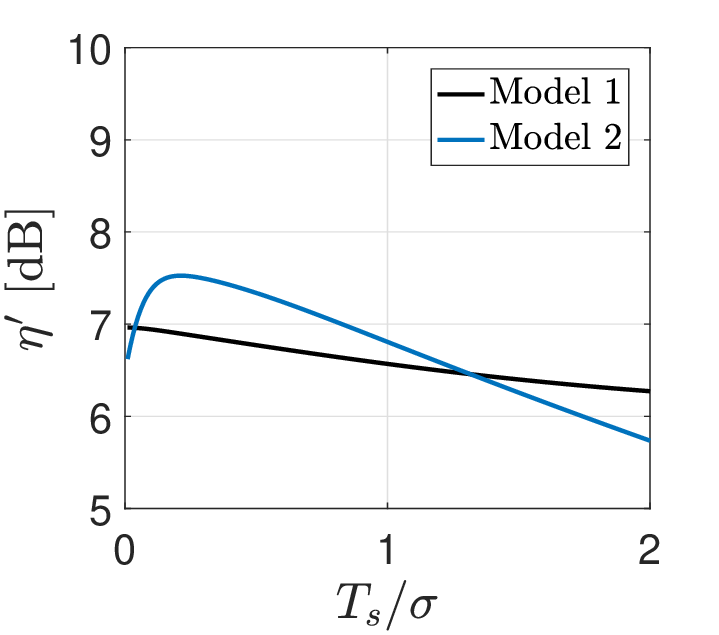}
\label{fig_Etap}}
\caption{Performance parameters introduced in Section \ref{Analysis} for conventional TR, calculated for both channel models as a function of the ratio between the symbol duration and the channel delay spread $T_s/\sigma$. Other parameters are: $L=33$, $L_1 \approx 8$, $L_2 \approx 17$, $\gamma = 0.4786$, $\sigma_1 = 8$ ns, and $\sigma_2=1.75\sigma_1$. (a) ISI power and usable power ratio, (b) Interference power and effective spatial focusing, and (c) apparent spatial focusing.}
\label{fig_UandEta}
\end{figure}

We analyze numerically the expressions found in the previous section for $\hat{U}$, $\hat{\eta}_{tr}$ and $\hat{\eta}'_{tr}$ in conventional TR. Fig. \ref{fig_UandEta} shows these results in terms of the ratio between the symbol time and the channel delay spread $T_s/\sigma$ (we use $T_s/\sigma_1$ for Model 2). We set the remaining parameters so they approximate Channel Model B in \cite{erceg2004} (i.e., $L=33$, $L_1 = 8$, $L_2 = 17$, $\gamma = 0.4786$ and $\sigma_2=1.75\sigma_1$). The number of antennas was set to $M = 4$. The ratio between the tap spacing $T_s$ and the delay spread parameter $\sigma$ determines the frequency selectivity of the channel: smaller values of $T_s/\sigma$ imply larger signal bandwidths or stronger scattering in the channel.

Fig. \ref{fig_U} shows that the usable power ratio for Model 1 $\hat{U}^{(1)}$ increases when the channel $T_s/\sigma$. However, the same behavior is not observed for Model 2, where the variations of $\hat{U}^{(2)}$ are not significant. Thus, no general conclusions on the ISI power behavior can be drawn, given its nonlinear dependence on several propagation parameters (see (\ref{eq_ISIpower21}) and (\ref{eq_ISIpower22})). Typical wideband channels, which are characterized by $T_s/\sigma < 1$, have a usable power ratio ranging from 5 dB to 15 dB in the simulated scenarios, which will limit the BER performance at high SNR.

Fig. \ref{fig_Eta} and Fig. \ref{fig_Etap} show the results for the effective spatial focusing and the total power focusing parameters. In both cases, an increase in the spatial focusing (beamforming capability) of conventional TR is observed for scenarios with stronger scattering and/or larger bandwidths (small $T_s/\sigma$). Also, $\hat{\eta}_{tr}>\hat{\eta}'_{tr}$ in all cases, which can be interpreted in the following way. Even though the received signal power at the desired user is between 6 dB and 8 dB (approximately) stronger than the signal power at the unintended receiver, an important fraction of these powers are composed of ISI. However, the usable power at the user's detector is actually significantly larger than the usable power at the unintended receiver (it can reach up to 25 dB in the simulated conditions). This is because the TR pre-filter is matched only to the desired user's CIR, and does not offer partial equalization at other spatial locations. It is also worth noting that an approximate upper bound on $\hat{\eta}_{eq}$ is the number of antennas (6 dB under the conditions described on Fig. \ref{fig_UandEta}) regardless of the channel model. We return to this issue later.

We also performed Monte Carlo simulations of the described conventional TR and ETR systems under tap separations of 2.5 ns, and 10 ns, consistent with current WLAN models as specified in \cite{erceg2004} and \cite{breit2009}. We calculated the performance parameters presented in Section \ref{Analysis} for 1000 channel realizations, with the transmission of $10^4$ frames of $10$ symbols in each one of them. The number of transmit antennas was $M = 4$ and the channel parameters were selected according to Table \ref{TableChannelModel}.

In concordance with the results in Fig. \ref{fig_UandEta}, the simulation shows that the total focusing performance improves by decreasing the tap separation, as presented in Table \ref{TableFocusing}. This is due to the increasing number of resolvable multipath components in the CIR, which are all coherently combined at the receiver thanks to the TR pre-filter. Also, the results are consistent with the closed form approximations (\ref{eq_focusingtr1}) and (\ref{eq_focusingtr2}). In the case of ETR, the approximate upperbound of 6 dB for the spatial focusing is satisfied in these scenarios, and a loss of between 1dB and 2dB is observed with respect to conventional TR. {This is caused by the effect of the equalizer over the desired signal and ISI power: larger delay spreads, smaller tap separations, or larger $L$ decrease the total received power under a constant $L_E$, as seen in Fig. \ref{fig_le}.} These results clearly demonstrate the potential of TR techniques for beamforming.

\begin{table*}[!t]
\caption{Spatial Focusing Performance Comparison}%
\label{TableFocusing}
\centering
\begin{tabular}{ccccccc}
\hline \hline
\multicolumn{1}{c}{\multirow{2}{*}{\centering{Tap spacing [ns]}}} & \multicolumn{2}{c}{Simulated $\eta'_{tr}$ [dB]} & \multicolumn{2}{c}{Theoretical $\hat{\eta}'_{tr}$ [dB]} & \multicolumn{2}{c}{Simulated $\eta'_{eq}$ [dB]} \\
 &  Model 1 & Model 2  & Model 1 & Model 2 & Model 1 & Model 2\\
 \hline
2.5  & 6.8 & 6.9 & 6.9 & 7.5 & 4.9 & 4.9\\
5  & 6.8 & 6.9 & 6.7 & 7.1 & 5.2 & 5.0\\
10 & 6.4 & 6.9 & 6.5 & 6.3 & 5.4 & 5.3\\
\hline \hline
\end{tabular}
\end{table*}

\subsection{BER Performance}

\begin{figure}[!t]
\centering
\hspace*{-10pt}\includegraphics[width=1.1\columnwidth]{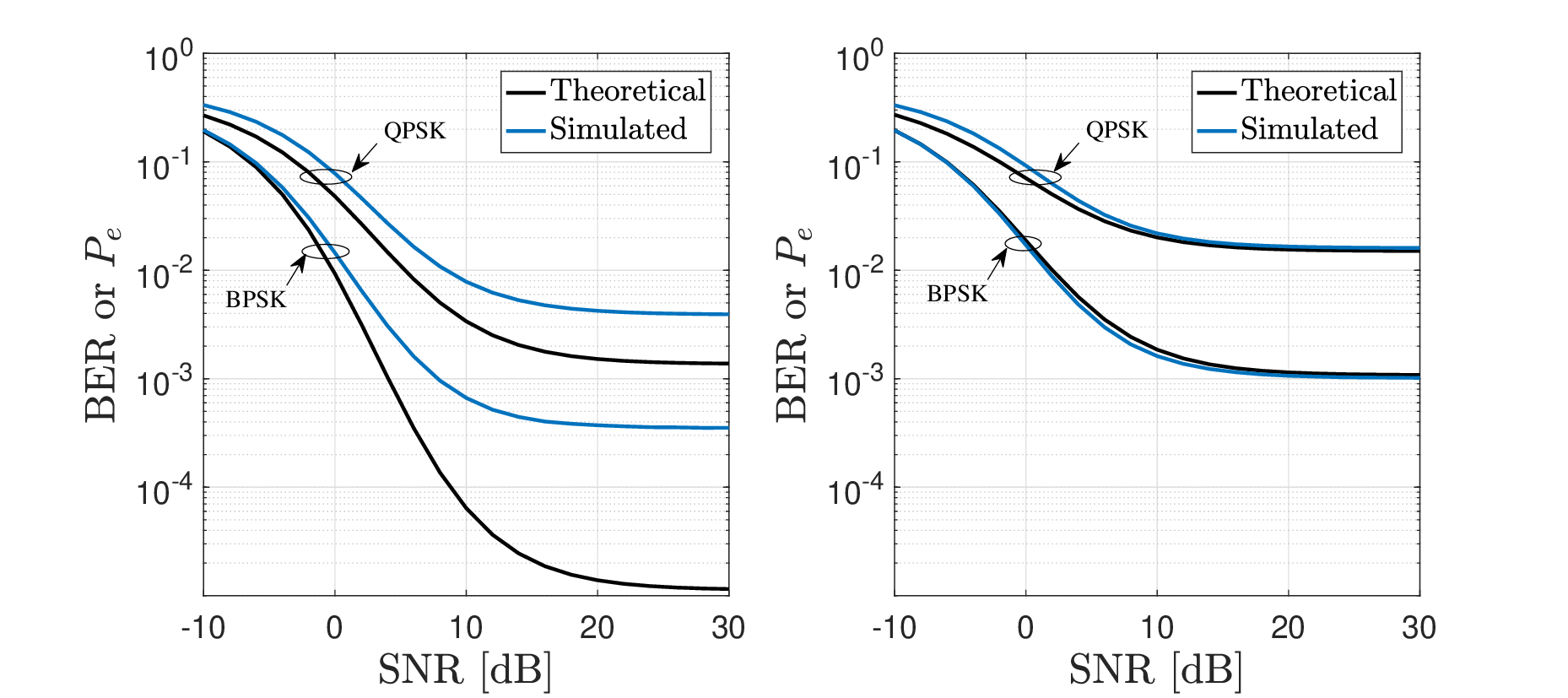}
\caption{Simulated BER and approximated probability of error as a function of the $SNR = \rho\Gamma/\sigma_z$ for BPSK. Simulation parameters are $T_s = 10$ ns, and $M = 4$ antennas. (left) Model 1, (right) Model 2.}
\label{fig_approximation}
\end{figure}

We calculated the BER of both conventional TR and ETR as a function of the signal to noise ratio defined as $SNR = \rho\Gamma/\sigma_z$. First, in Fig. \ref{fig_approximation} we verify our approximation to the probability of error in conventional TR using the closed form expressions (\ref{eq_signalpower}), (\ref{eq_ISIpower21}) and (\ref{eq_ISIpower22}). BPSK and QPSK modulations were used. The difference between our theoretical $P_e$ approximation and the simulated BER results improves in Model 2. This is due to the smaller variance of the normalization factor in Model 2, as explained in Appendix \ref{AppB}. It is worth noting that the approximation accuracy to the probability of error is highly dependent on the specific set of parameters describing the propagation conditions.

It is observed that Model 1 (weaker scattering) has a better BER performance, as expected from the usable power ratio results in Fig. \ref{fig_U}. In addition, it is clear that the BER in both modulations is too high to be of practical use in the scenarios considered here; this is because the ISI power causes a lower bound on the probability of bit error, as stated in Section \ref{Analysis}. Thus, the relevance of the proposed ETR technique to overcome this problem is evident.

%\begin{figure}[!t]
%\centering
%\subfloat[]{\includegraphics[width=0.5\columnwidth]{Pe801.eps}
%\label{fig_801}}
%\subfloat[]{\includegraphics[width=0.5\columnwidth]{Pe802.eps}
%\label{fig_802}}
%\newline
%\centering
%\subfloat[]{\includegraphics[width=0.5\columnwidth]{Pe1601.eps}
%\label{fig_1601}}
%\subfloat[]{\includegraphics[width=0.5\columnwidth]{Pe1602.eps}
%\label{fig_1602}}
%\newline
%\caption{Simulated BER as a function of the $SNR = \rho\Gamma$ for BPSK modulation. (a) $T_s=5$ ns Model 1, (b) $T_s=5$ ns Model 2, (c) $T_s=2.5$ ns Model 1, (d) $T_s=2.5$ ns Model 2.}
%\label{fig_approximation}
%\end{figure}

Fig. \ref{fig_tretr} shows the simulated BER performance for the TR and ETR and the lower bound for the probability of error using BPSK. The number of antennas is $M = 4$. The equalizer's length is the same as the CIR length, i.e. $L_E=L$. Again it is noted that conventional TR has a lower bound on the BER caused by ISI, and that the performance deteriorates by increasing the delay spread (Model 2) or the bandwidth due to stronger ISI power. Variations of the BER in ETR are not significant with respect to changes in model parameters. ETR outperforms conventional TR under any SNR by mitigating the ISI, so its BER performance approaches that of the AWGN channel.

\begin{figure}[t]
\centering
\hspace*{-10pt}\includegraphics[width=1.1\columnwidth]{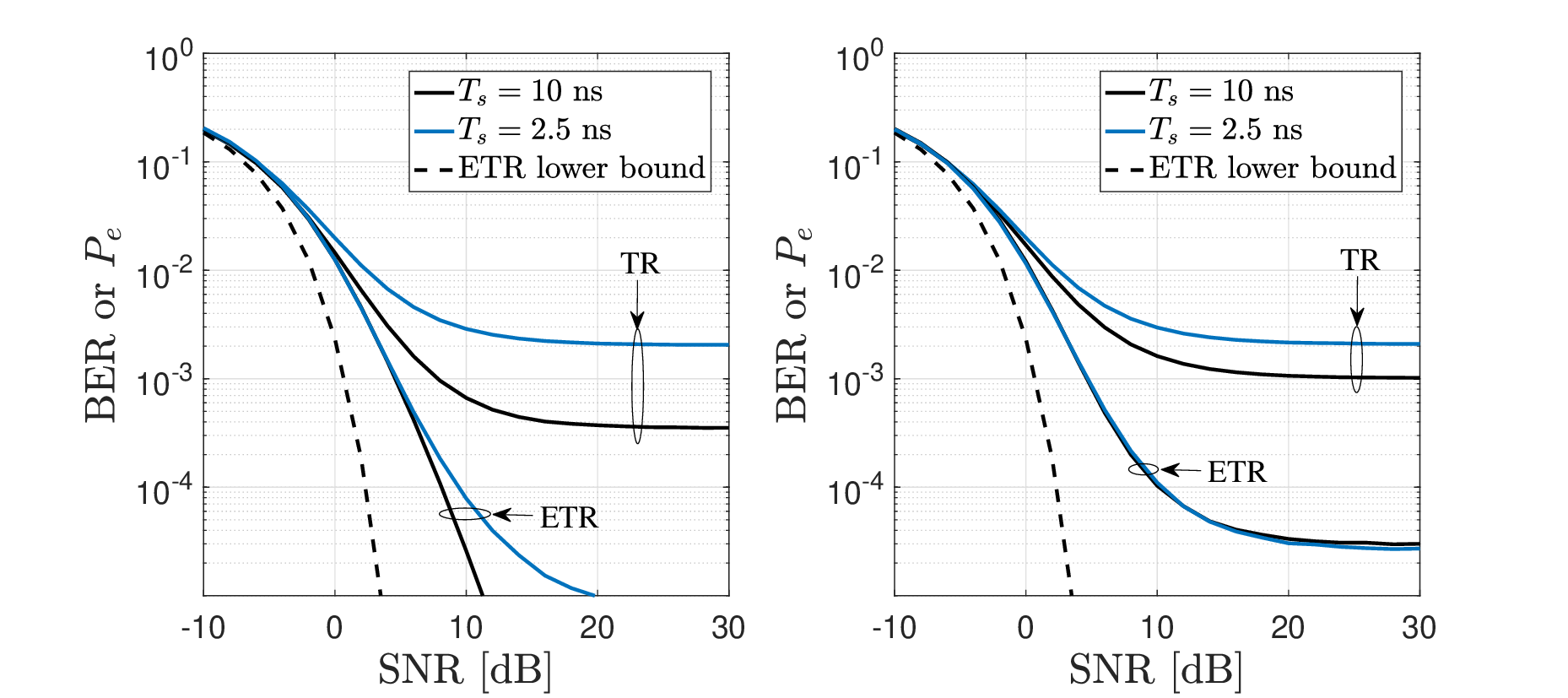}
\caption{Simulated BER as a function of the $SNR = \rho\Gamma/\sigma_z$ for BPSK. Comparison between conventional TR and ETR with different symbol durations. (left) Model 1, (right) Model 2.}
\label{fig_tretr}
\end{figure}

Fig. \ref{fig_antennas} shows the simulated BER of conventional TR and ETR under different channel models, and number of antennas. BER performance variations in ETR is are significant with respect to the channel models. However, when increasing the number of antennas from 4 to 8, conventional TR performance improves significantly, approaching that of ETR. Specifically, lower BER can be achieved at high SNR when increasing the number of antennas due to the linear dependence of the usable power ratio on $M$ (as seen in (\ref{eq_UsablePower1}) and (\ref{eq_UsablePower2})). This phenomena can be harnessed in systems with a large number of transmit antennas, where the equalization properties of TR can allow sufficiently low BER without further processing \cite{viteri2015}. 

\begin{figure}[t]
\centering
\hspace*{-10pt}\includegraphics[width=1.1\columnwidth]{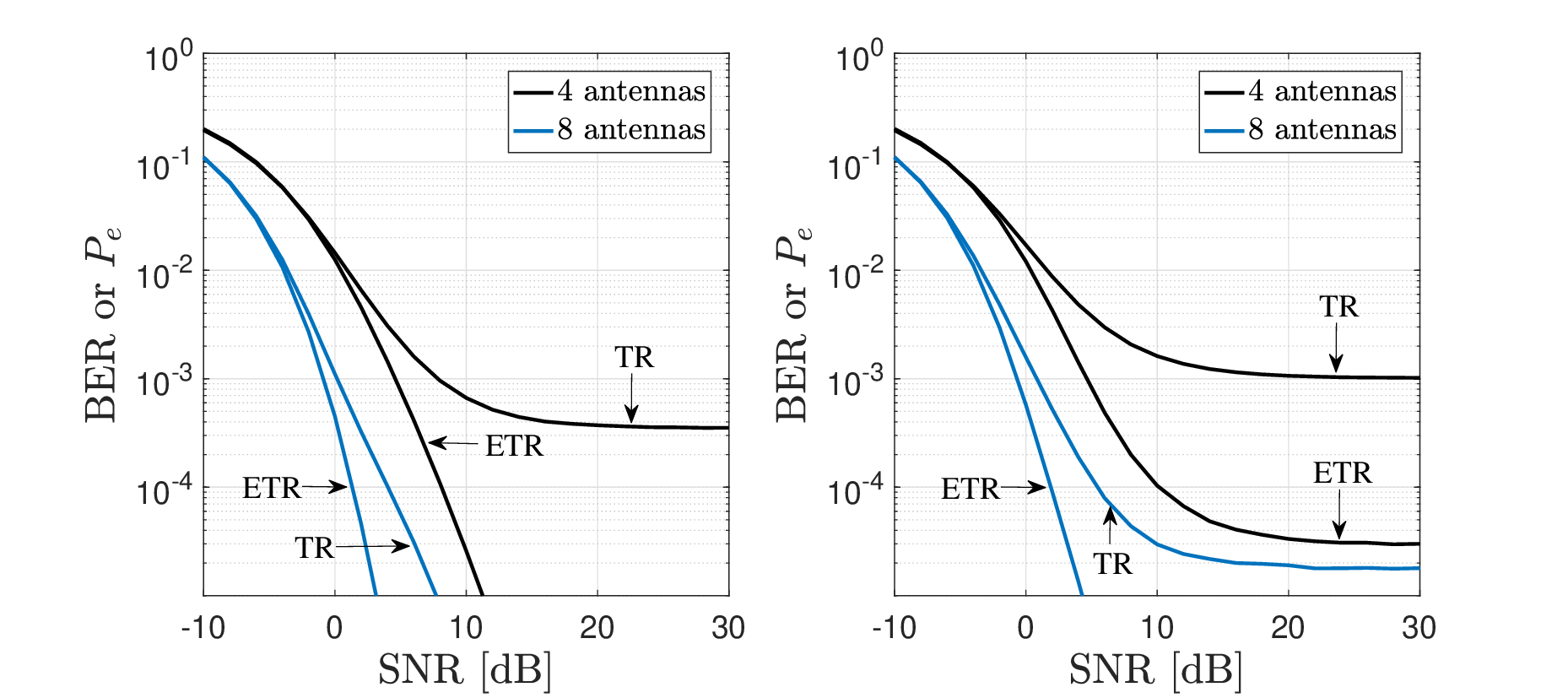}
\caption{Simulated BER as a function of the $SNR = \rho\Gamma/\sigma_z$ for BPSK. Comparison between conventional TR and ETR accross different channel models and number of antennas ($M$). (left) Channel 1, (right) Channel 2.}
\label{fig_antennas}
\end{figure}

\section{Conclusions}
\label{Conslusions}

We have analyzed a baseband TR beamforming system using two propagation models commonly used in indoor wireless communications. In particular, this analysis is relevant for pico and femtocells in conventional wideband systems such as WiFi networks. We derived a novel closed-form approximation for the ISI power in such scenarios in single-user wideband systems without rate back-off, and compare the probability of bit error obtained under different propagation conditions.

We analyzed parameters for the spatial focusing and time compression properties of TR beamforming and found closed-form approximations for them. By analyzing this parameters under two channel models, we found that TR performance is strongly dependent on propagation conditions. Specifically, there are significant variations on the ISI power depending on the power delay profile, the symbol duration (bandwidth), and the channel delay spread. Hence, no general conclusions can be extracted for the ISI for specific propagation conditions.

We then proposed an equalized TR technique as a solution to mitigate ISI. ETR uses a single ZF pre-equalizer at the transmitter in cascade configuration with the TR pre-filters. Unlike previous approaches, we analytically showed that the proposed technique greatly enhances the performance of conventional TR with low impact to its beamforming capability. An upper bound on the received power of ETR was also derived, which corresponds to a lower bound on the probability of bit error.

The spatial focusing performance of conventional TR and ETR was analyzed by calculating the signal power at an unintended receiver with uncorrelated CIR. We showed that the effective power ratio and the apparent power ratio between the receiver and the interfered user increase with either the channel delay spread or the signal bandwidth. Moreover, it was shown that the use of ETR has a small impact over this spatial focusing parameters.

By means of numerical simulations, we verified that the proposed ETR technique outperforms conventional TR with respect to the BER under any SNR, even though the total received power is greater for conventional TR. We also verified the accuracy of the approximation to the ISI power in conventional TR and found that it improves for channel model with stronger delayed components (Model 2 defined here).

\appendices
\section{Upper Bound on the ETR Received Power}
\label{AppA}
In this Appendix, we derive an upper bound for the received power with the proposed ETR technique. From (\ref{eq_eqnormal}), (\ref{eq_equalizerfreq}) and Parseval's theorem, it follows that
\begin{IEEEeqnarray}{rCl}
P_g & = & \frac{1}{2L+L_E-2} \sum_{i=1}^{M_T} \sum_{k=0}^{2L+L_E-3} \left| H_i^*[k] G_{zf}[k] e^{-j\frac{2\pi(L-1)}{2L+L_E-2}k} \right|^2 \nonumber \\
	& = & \frac{1}{2L+L_E-2} \sum_{i=1}^{M_T} \sum_{k=0}^{2L+L_E-3} \left| H_i[k] \right|^2 \frac{1} {\left| \sum_{p=1}^{M_T} \left| H_p[k] \right|^2 \right|^2}  \nonumber \\
	& = & \frac{1}{2L+L_E-2} \sum_{k=0}^{2L+L_E-3} \frac{1} { \sum_{i=1}^{M_T} \left| H_i[k] \right|^2 }, \nonumber
\end{IEEEeqnarray}
where we have used zero padding in order to represent the linear convolution. Now, the received signal power using ETR beamforming is
\begin{IEEEeqnarray}{rCl}
\label{eq_receivedapp}
\hspace*{-20pt} P_{eq} & = & \rho \, \mathbb{E} \left[ \frac{1}{P_g} \right] =  \rho \, \mathbb{E} \left[ \frac{2L+L_E-2}{\sum_{k=0}^{2L+L_E-3} \frac{1} { \sum_{i=1}^{M_T} \left| H_i[k] \right|^2 }}  \right].
\end{IEEEeqnarray}
Note that the expression inside the expectation operator in (\ref{eq_receivedapp}) is a concave function of $\left| H_i[k] \right|^2$ (i.e. it is a double composition of an affine function and its reciprocal) \cite[Sec. 3.2.4]{boyd2004}. Hence, from Jensen's inequality we get
\begin{IEEEeqnarray}{rCl}
\label{eq_receivedappbound}
P_{eq} & \leq & \frac{\rho \, (2L+L_E-2)}{\sum_{k=0}^{2L+L_E-3} \frac{1} { \sum_{i=1}^{M_T} \mathbb{E} \left[ \left| H_i[k] \right|^2 \right] }}.
\end{IEEEeqnarray}
By using (\ref{eq_channelnormalization}), uncorrelated scattering and the DFT definition, we also have
\begin{IEEEeqnarray}{rCl}
\label{eq_receivedappbound2}
\mathbb{E} \left[ \left| H_i[k] \right|^2 \right] & = & \mathbb{E} \left[ \sum_{m=0}^{2L+L_E-3} \sum_{n=0}^{2L+L_E-3} h_i[m] h_i^*[n] \right. \nonumber \\
												  & & \qquad \qquad \qquad \times \ e^{-j\frac{2 \pi m}{2L+L_E-2}k} e^{j\frac{2 \pi n}{2L+L_E-2}k} \Bigg] \nonumber \\
& = &  \mathbb{E} \left[ \sum_{n=0}^{2L+L_E-3} \left| h_i[n] \right|^2 \right] = \Gamma.
\end{IEEEeqnarray}
Replacing (\ref{eq_receivedappbound2}) in (\ref{eq_receivedappbound}):
\begin{IEEEeqnarray}{rCl}
\label{eq_receivedappbound3}
P_{eq} & \leq & \rho \, M \, \Gamma. \nonumber
\end{IEEEeqnarray}

\section{Approximation to the ISI power in conventional TR}
\label{AppB}
In this appendix, we derive an approximation to the ISI power in conventional TR systems and analyze the approximation error using the variance of the normalization factor. From (\ref{eq_channelpower}) and (\ref{eq_ISIpower}), the ISI power is given by:
\begin{IEEEeqnarray}{rCl}
\label{eq_ISIpowera1}
\hspace*{-18pt} P_{ISI} & = & \rho \, \mathbb{E} \left[ \frac{\left| \sum\limits_{i=1}^{M} \sum\limits_{\substack{ l=0 \\ l \neq L-1}}^{2L-2} \sum\limits_{n=0}^{L-1}h_i[n]h_i^*[L-1-l+n] \right|^2}{\sum\limits_{i=1}^{M} \sum\limits_{l=0}^{L-1}|h_i[l]|^2}  \right].
\end{IEEEeqnarray}
Let $a$ and $b$ be two correlated random variables. According to \cite{rice2008} and \cite{rice2009}, an expansion for the expectation of the ratio of $a$ and $b$ is
\begin{IEEEeqnarray}{rCl}
\label{eq_approximation1}
\hspace*{-20pt} \mathbb{E}\left[\frac{a}{b} \vert b \neq 0 \right] =  \frac{\mathbb{E}\left[a\right]}{\mathbb{E}\left[b\right]} + \sum_{i=1}^\infty (-1)^i \frac{\mathbb{E}\left[  a \right] \langle ^i b \rangle + \langle a, ^i b \rangle}{\mathbb{E}\left[b\right]^{i+1}},
\end{IEEEeqnarray}
where $\langle ^i b \rangle = \mathbb{E}\left[(b-\mathbb{E}[b])^i\right]$ is the $i$-th central moment of $b$ and $\langle a , ^i b \rangle = \mathbb{E}\left[(a-\mathbb{E}[a])(b-\mathbb{E}[b])^i\right]$ is the $i$-th mixed central moment of $b$ and $a$. Thus, if we only consider the first term in the expansion, we can approximate (\ref{eq_ISIpowera1}) as
\begin{IEEEeqnarray}{rCl}
\label{eq_ISIpowera2}
\hspace*{-18pt} \hat{P}_{ISI} & = & \rho \,  \frac{\mathbb{E} \left[\left| \sum\limits_{i=1}^{M} \sum\limits_{\substack{ l=0 \\ l \neq L-1}}^{2L-2} \sum\limits_{n=0}^{L-1}h_i[n]h_i^*[L-1-l+n] \right|^2 \right]}{\mathbb{E} \left[\sum\limits_{i=1}^{M} \sum\limits_{l=0}^{L-1}|h_i[l]|^2\right]}.
\end{IEEEeqnarray}
We will analyze the approximation error later in this section. Since the channel has uncorrelated scattering and $h_i[n]$ has zero mean, we note that
\begin{IEEEeqnarray}{rCl}
\label{eq_ISI1}
&& \mathbb{E} \left[ h_i[n]h_i^*[L-1-l+n]h_{i'}^*[n']h_{i'}[L-1-l'+n'] \right] = 0 \nonumber \\
&&\text{if } i \neq i' \text{ or } n \neq n'.\nonumber
\end{IEEEeqnarray}
Thus, the only non-zero terms in the numerator of (\ref{eq_ISIpowera2}) are of the form \linebreak
$\mathbb{E} \left[ \left| h_i[n] \right|^2 \left| h_i[L-1-l+n] \right|^2 \right]$. Using these results we get:
\begin{IEEEeqnarray}{lCl}
\label{eq_ISIpowera3}
\hspace*{-20pt} \hat{P}_{ISI} & = & \frac{\rho}{M \Gamma} \sum_{\substack{ l=0 \\ l \neq L-1}}^{2L-2} \sum_{i=1}^{M} \sum_{\substack{ n=0 \\ n \leq l \\ n \geq l-L+1 }}^{L-1} \mathbb{E} \left[ \left| h_i[n] \right|^2 \right] \nonumber \\
&& \qquad \qquad \qquad \qquad \times \  \mathbb{E} \left[ \left| h_i[L-1-l+n] \right|^2 \right], 
\end{IEEEeqnarray}
where the constraints in the sum over $n$ come from the definition of the PDP for $n\in\{0,\ldots,L-1\}$, so $0 \leq L-1-l+n \leq L-1$ must hold. Now, notice that if we only consider the first term in the expansion (\ref{eq_approximation1}), the equality holds if the variance of the denominator is vanishingly small. Thus, we use this variance (denoted $\text{Var}[P_h]$) as a measure of the approximation error. Using the given channel models, this variance is given in (\ref{eq_varph1}) and (\ref{eq_varph1}).
\begin{figure}[!t]
\centering
\includegraphics[width=0.7\columnwidth]{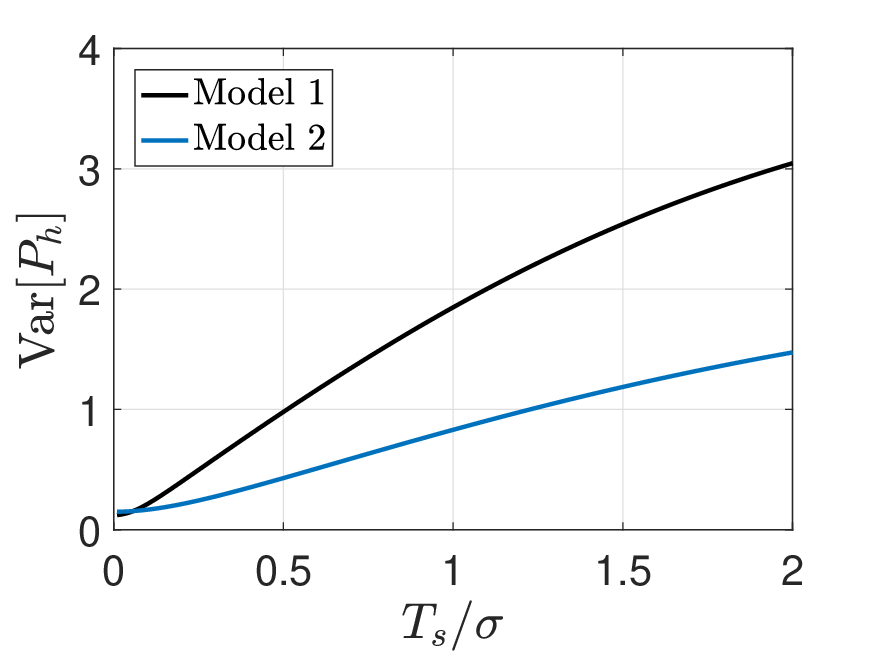}
\caption{Variance of the normalization factor for Model 1 - eq. (\ref{eq_varph1}), and Model 2 - eq. (\ref{eq_varph2}). Note that the variance is diminishingly small when the ratio $T_s/\sigma$ is small (richer scattering or large bandwidth). Thus, a smaller approximation error between $P_{ISI}$ and $\hat{P}_{ISI}$ is expected in Model 2, and also for smaller tap separations and/or larger delay spreads.  Other parameters are: $L=33$, $L_1 \approx 8$, $L_2 \approx 17$, $\gamma = 0.4786$, $\sigma_1 = 8$ ns, and $\sigma_2=1.75\sigma_1$.}
\label{fig_varpisi}
\end{figure}
\begin{IEEEeqnarray}{rCl}
\label{eq_varph1}
\text{Var}[P_h]^{(1)} & = & M \, \Gamma^2 \frac{\left( 1 + e^{-\frac{LT_s}{\sigma}} \right)\left( 1 - e^{-\frac{T_s}{\sigma}} \right)}{\left( 1 + e^{-\frac{T_s}{\sigma}} \right)\left( 1 - e^{-\frac{LT_s}{\sigma}} \right)},
\end{IEEEeqnarray}
\begin{figure*}[t]
\begin{IEEEeqnarray}{rCl}
\label{eq_varph2}
\text{Var}[P_h]^{(2)} & = & M \, \Gamma^2 \, \frac{ \left( \sum\limits_{n=0}^{L_2-1} e^{-\frac{2nT_s}{\sigma_1}} + \gamma^2 \sum\limits_{n=L_1}^{L-1} e^{-\frac{2(n-L_1)T_s}{\sigma_2}}  + 2 \gamma \sum\limits_{n=L_1}^{L_2-1} e^{-\frac{nT_S}{\sigma_1}} e^{-\frac{(n-L_1)T_S}{\sigma_2}} \right)}{\left( \sum\limits_{n=0}^{L_2-1} e^{-\frac{nT_s}{\sigma_1}} + \gamma \sum\limits_{n=L_1}^{L-1} e^{-\frac{(n-L_1)T_s}{\sigma_2}}\right)^2}.
\end{IEEEeqnarray}
\hrulefill
\end{figure*}
\setcounter{equation}{31}

Fig. \ref{fig_varpisi} shows the variance of the normalization factor as a function of the CIR length for different values of $T_s/\sigma$ or $T_s/\sigma_1$, and according to the channel model. The variance approaches zero for decreasing values of $T_s/\sigma$. This means that the approximation given by (\ref{eq_ISIpowera3}) improves in scenarios with larger delay spread, or smaller tap separation. For example, it is observed that a better approximation is achieved for Model 2 due to the stronger scattering (delayed components with larger power) under same CIR length.

\section{Total Power at an Unintended Receiver using ETR}
\label{AppC}
In this section we derive a closed-form approximation for the total power at an unintended receiver when our proposed ETR technique is used. For this, we use the same procedure as in \ref{AppB}. From (\ref{eq_intreceived}), the total power at an unintended receiver is
\begin{IEEEeqnarray}{lCl}
P_{int}^{eq}  & = & \rho \, \mathbb{E} \left[ \frac{ \sum\limits_{n=0}^{2L+L_E-3} \left| \sum\limits_{i=1}^{M_T} g[n] \otimes h_i^*[L-1-n] \otimes h_{u,i}[n] \right|^2}{\sum\limits_{n=0}^{2L+L_E-3} \sum\limits_{i=1}^{M_T} \left|  g[n] \otimes h_i^*[L-1-n] \right|^2} \right], \nonumber \\
\end{IEEEeqnarray}
which takes into account the desired signal power and the ISI power. Using Parseval's theorem,
\begin{IEEEeqnarray}{lCl}
\label{eq_app1}
P_{int}^{eq}  & = & \rho \, \mathbb{E} \left[ \frac{ \sum\limits_{k=0}^{2L+L_E-3} \left| \sum\limits_{i=1}^{M_T} G[k] H_i^*[k] H_{u,i}[k] e^{-j\frac{2\pi(L-1)}{2L+L_E-2}k} \right|^2}{\sum\limits_{k=0}^{2L+L_E-3} \sum\limits_{i=1}^{M_T} \left|  G[k] \right|^2 \left|   H_i[k] \right|^2} \right],  \nonumber \\
\end{IEEEeqnarray}
where $H_{u,i}[k]$ is the DFT of $h_{u,i}[n]$. Using the same expansion as in \ref{AppB}, (\ref{eq_app1}) can be approximated as
\begin{IEEEeqnarray}{lCl}
\label{eq_pint1}
\hat{P}_{int}^{eq} & = & \rho \frac{ \mathbb{E} \left[\sum\limits_{k=0}^{2L+L_E-3} \left| \sum\limits_{i=1}^{M_T} G[k] H_i^*[k] H_{u,i}[k] e^{-j\frac{2\pi(L-1)}{2L+L_E-2}k} \right|^2 \right]}{ \mathbb{E} \left[ \sum\limits_{k=0}^{2L+L_E-3} \sum\limits_{i=1}^{M_T} \left|  G[k] \right|^2 \left|   H_i[k] \right|^2 \right]}, \nonumber \\
\end{IEEEeqnarray}
where $H_i[k]$ and $H_{u,i'}[k']$ have zero mean and are uncorrelated for all $i$, $k$, $i'$ and $k'$. Also, $H_{u,i}[k]$ and $H_{u,i}[k]$ are uncorrelated $\forall k$ if $i\neq i'$. Then, (\ref{eq_pint1}) becomes
\begin{IEEEeqnarray}{lCl}
\label{eq_pint2a}
\hat{P}_{int}^{eq} & = & \rho \frac{ \mathbb{E} \left[\sum\limits_{k=0}^{2L+L_E-3} \sum\limits_{i=1}^{M_T} \left| G[k] \right|^2  \left| H_i[k] \right|^2 \left| H_{u,i}[k] \right|^2 \right]}{ \mathbb{E} \left[ \sum\limits_{k=0}^{2L+L_E-3} \sum\limits_{i=1}^{M_T} \left|  G[k] \right|^2 \left|   H_i[k] \right|^2 \right]} \nonumber \\
& = & \rho \frac{ \sum\limits_{k=0}^{2L+L_E-3} \sum\limits_{i=1}^{M_T} \mathbb{E} \left[ \left| G[k] \right|^2  \left| H_i[k] \right|^2 \right] \mathbb{E} \left[ \left| H_{u,i}[k] \right|^2 \right]}{  \sum\limits_{k=0}^{2L+L_E-3} \sum\limits_{i=1}^{M_T} \mathbb{E} \left[  \left|  G[k] \right|^2 \left|   H_i[k] \right|^2 \right]}. \nonumber 
\end{IEEEeqnarray}
Finally, from the channel power normalization, $\mathbb{E} \left[ \left| H_{u,i}[k] \right|^2 \right] = \Gamma$, so
\begin{IEEEeqnarray}{lCl}
\label{eq_pint3}
\hat{P}_{int}^{eq} & = & \rho \, \Gamma. \nonumber
\end{IEEEeqnarray}

%\section*{Acknowledgement}
%This work has been supported in part by the Ohio Supercomputer Center (OSC) under grants PAS-0061 and PAS-0110, and by a Fulbright Colombia Fellowship.

\bibliographystyle{IEEEtran}
\IEEEtriggeratref{27}
\bibliography{TR-ETR}

\end{document}